\documentclass[a4paper]{article}

\usepackage{latexsym}

\newcommand{\ov}{\overline}

\newcommand{\p}{\partial}

\newcommand{\wt}{\widetilde}

\newcommand{\href}[1]{{\ref{#1}}}
\newcommand{\hlabel}[1]{\label{#1} 
  }

\newcommand{\hcite}[1]{
 \cite{#1}}
\newcommand{\hbibitem}[1]{\bibitem{#1} 
  }
\newcommand{\hepth}[1]{{hep-th/#1}}

\newcommand{\eq}{\end{equation}}
\newcommand{\be}[4]{ 
\begin{equation} 
\hlabel{#1#2#3#4}}  
\newcommand{\bea}[4]{ 
\begin{eqnarray}
\hlabel{#1#2#3#4}}
\newcommand{\eqa} {\end{eqnarray}}

\newcommand{\eeqa}{\end{eqnarray}}
\newcommand{\ee}{\end{equation}}
\newcommand{\ba}{\begin{array}}
\newcommand{\ea}{\end{array}}

\def\C{ {\cal C}} 
\newcommand{\G}{\Gamma} 
\def\Kahler{K\"{a}hler}

\newcommand{\z}{\zeta} 
\newcommand{\eeq}{\end{equation}}
\newcommand{\beq}{\begin{equation}}

\def\hybrid{\topmargin -20pt  
      \oddsidemargin 0pt
      \headheight 0pt   
      \headsep 0pt
      \textwidth 6.25in 
      \textheight 9.5in 
      \marginparwidth .875in
      \parskip 5pt plus 1pt   \jot = 1.5ex}

\hybrid

\let\LARGE=\large

\let\large=\normalsize

\begin{document}

\thispagestyle{empty}
\rightline{HUB-EP-97/34}
\rightline{UPR-754-T}
\rightline{hep-th/9706096}
\rightline{June 1997}
\vspace{1truecm}

\begin{center}
{\LARGE \bf {From} Type IIA Black Holes to T-dual Type IIB D-Instantons\\
in $N=2$, $D=4$ Supergravity}
\end{center}

\vspace{1.2truecm}
\begin{center}
{\bf Klaus Behrndt$^a$, Ingo Gaida$^b$, Dieter L\"ust$^a$, 
Swapna Mahapatra$^a$,
Thomas Mohaupt$^a$}
\footnote{E-mail: 
behrndt@qft2.physik.hu-berlin.de, gaida@cvetic.hep.upenn.edu, 
luest@qft1.physik.hu-berlin.de, \hfill \\ \hspace*{1.5cm} 
swapna@qft2.physik.hu-berlin.de, mohaupt@qft2.physik.hu-berlin.de}
\end{center}

\begin{center}
$^a${\em Humboldt-Universit\"at zu Berlin,
Institut f\"ur Physik, \\ 
D-10115 Berlin, Germany}\\
$^b$ {\em Department of Physics and Astronomy,
University of Pennsylvania, \\ Philadelphia, PA 19104-6396, U.S.A.}
\end{center}

\vspace{1.2truecm}



\begin{abstract}

\noindent
We discuss the T-duality between the 
solutions of type IIA versus IIB superstrings compactified on
Calabi-Yau threefolds. Within the context of the $N=2$, $D=4$
supergravity effective Lagrangian, the T-duality transformation
is equivalently described by the c-map, which relates the special
K\"ahler moduli space of the IIA $N=2$ vector multiplets to the
quaternionic moduli space of the $N=2$ hyper multiplets on the type IIB
side (and vice versa). Hence the T-duality, or 
c-map respectively, transforms the
IIA black hole   solutions, originating from even dimensional IIA branes,
of the special K\"ahler effective action, into IIB D-instanton
solutions of the
IIB quaternionic
$\sigma$-model action, where
the D-instantons can be obtained by compactifying  odd IIB D-branes
on the internal Calabi-Yau space.
We construct via this mapping a broad class
of D-instanton solutions in four dimensions  which are
determinded by  a set 
of harmonic functions plus the underlying topological Calabi-Yau data.

\end{abstract}

\bigskip \bigskip
\newpage

\section{Introduction}

Much of the recent progress in non-perturbative string and
field theory is due to the insight that open strings with
Dirichlet boundary conditions, which had been discovered
already some time ago \hcite{DaiLeiPol},
can be used to describe type I and 
type II string theory in non-perturbative, solitonic or instanton 
backgrounds \hcite{Pol}.
Specifically, the Dirichlet (D) $p$-branes, carrying the charges
of the Ramond-Ramond (RR) gauge potentials,  
arise as the solutions
of type IIA supergravity in ten dimensions for $p$ even, whereas the 
Dirichlet
$p$-branes, with $p$ odd, appear as the solitons in the type IIB superstring.
One of the most important results in this
context, which is obtained by
counting D-brane states, is the computation \hcite{StroVa}
of the statistical
entropy of five- and four-dimensional extremal black holes
in type II strings, which are
obtained by wrapping certain configurations of higher dimensional IIA
or M-theory branes
around the internal cycles of the compactification space. This statistical
entropy agrees with the macroscopic,
Bekenstein-Hawking entropy which is proportional
to the finite horizon of the extremal black hole solutions.

Recently stationary extremal solutions of 
four-dimensional $N=2$ supergravity coupled
to  $N=2$ vector multiplets were investigated in some detail \hcite{be/lu}.
Since the $N=2$ vector couplings are determined by special K\"ahler
geometry \hcite{SpecK1,special_K}, it follows naturally that the stationary
$N=2$ background fields, like the  metric 
or moduli fields,  can be expressed entirely by objects which appear
in $N=2$ special geometry, like the K\"ahler potential or the
$U(1)$ K\"ahler connection; the corresponding
solutions of the field equations are completely determined by a set of
harmonic functions. 
Of particular interest are the $N=2$ extremal black hole solutions.
In the type IIA string compactifications on a Calabi-Yau space, $N=2$
black holes can be obtained by wrapping three ten-dimensional 4-branes
around the internal Calabi-Yau 4-cycles plus adding also a ten-dimensional
0-brane which lives in the internal space.
The macroscopic Bekenstein Hawking entropy of the $N=2$ 
black holes follows from the
extremization  of the $N=2$ central charge of the underlying $N=2$ 
super-algebra
with respect to the moduli fields \hcite{FeKaStr}. By means of this
procedure it is possible to compute the classical entropy of static
black holes \hcite{classical} as well as to incoporate quantum 
corrections 
\hcite{quantumcor}. So the full entropy depends in the  
heterotic string framework on
space-time instanton numbers, whereas in the type IIA compactifications
on Calabi-Yau threefolds the entropy depends on the topological data
of the Calabi-Yau space, like the intersection numbers, the Euler number
and the rational world-sheet instanton  numbers.
In addition, it was shown in \hcite{BeMo} that the contribution of the
intersection numbers to the macroscopic entropy matches the
microscopic entropy obtained by considering one internal IIA 0-brane plus
three IIA 4-branes, wrapped around the internal 
Calabi-Yau 4-cycles.

An interesting class of  non-perturbative solutions 
of type IIB supergravity, which break half of the
supersymmetries, are given
by the ten-dimensional D-instantons \hcite{GibGrePer}, 
i.e. $(-1)$-branes in Euclidean space. 
They are dual to the type IIB 7-branes, which have to be included in
the F-theory interpretation \hcite{Vaf} of the type IIB superstring.
In addition, D-instantons were shown \hcite{GreenVanhove}
to lead to non-perturbative
corrections to some higher order curvature terms in the effective
gravitational action of type IIB string compactifications.

T-duality between the type IIA and the type IIB string (on a circle)
transforms the even and odd $p$-configurations into each other 
\hcite{DaiLeiPol,BerDeR}.
Hence by T-duality the D-instanton of IIB is related to the
0-brane solution of IIA, i.e to a black hole.
In this paper we will investigate the T-duality between the
supersymmetric 
solitonic solutions of the IIA and IIB theories, compactified
on the same Calabi-Yau threefold. In particular, we will map the
four-dimensional $N=2$ IIA black holes on four-dimensional D-instanton
solutions 
of the type IIB $N=2$ effective actions. 
{From} the ten-dimensional $p$-brane point of view, 
the T-duality transforms the
$(0,4,4,4)$ IIA black hole configuration into the $(-1,3,3,3)$ IIB D-instanton
solution, whereas both brane configurations preserve 1/8 of the ten-dimensional
supersymmetries, i.e. half of the four-dimensional $N=2$ supersymmetries
of the compactified type II strings.

The T-duality with respect to the time coordinate
among the four-dimensional $N=2$ IIA  and IIB superstrings,
compactified on the same Calabi-Yau space, is completely equivalent to 
the $N=2$ c-map between these two theories, which was considered some
time ago in \hcite{cec,ferr}.
Namely, comparing the IIA and IIB spectra on a Calabi-Yau space
with Hodge numbers $h_{1,1}$ and $h_{2,1}$, one realizes that the
$c$-map (almost) exchanges the number of $N=2$ vector and $N=2$ hyper
multiplets: $N_V^{(A)}=h_{1,1}\leftrightarrow N_H^{(B)}-1=h_{1,1}$,
$N_H^{(A)}-1=h_{2,1}\leftrightarrow N_V^{(B)}=h_{2,1}$.
More specifically, the $N_V$ electric/magnetic
Abelian gauge fields in the RR sector of the
IIA superstring get transformed by T-duality
 into complex scalars in the RR sector of the type IIB theories, which
are then members of $N=2$ hyper multiplets. Moreover the universal IIA
graviphoton vector field gets mapped into the universal RR scalar field
which, together with the NS-NS dilaton-axion field, builds the univeral IIB
hyper multiplet. So the T-duality transformations relate the special
K\"ahler moduli space $K_{h_{1,1}}^{(A)}$
of complex dimension $h_{1,1}$ of the IIA vector  multiplets 
to the quaternionic moduli space $Q_{h_{1,1}+1}^{(B)}$ of
quaternionic dimension $h_{1,1}+1$ of the hyper multiplets (and vice versa
for the IIA hyper multiplets and the IIB vector multiplets).
Hence, whereas the IIA black holes are the solutions of the IIA special
K\"ahler Lagrangian, the T-dual IIB D-instantons arise as the
solutions of the field equations of the IIB quaternionic non-linear
$\sigma$-model.
The aim of this paper is to explore this relation between the solutions
of special K\"ahler IIA geometry and the solutions of the quaternionic
geometry on the type IIB side. Applying the T-duality or c-map
on known type IIA black hole solutions we will find a full new class
of type IIB D-instanton type of solutions. 
In the simplest case, the Reissner-Nordstrom charged
black-hole, which arises as the solution of the
universal type IIA Lagrangian of the $N=2$ graviphoton field,
gets mapped to the D-instanton solution which solves
the field equations of the quaternionic $\sigma$-model
with the universal dilaton hyper multiplet field. 
Via T-duality the IIA black hole metric background  gets
transformed into a non-trivial background for the NS-NS dilaton-axion field
on the type IIB side. More generally, we will consider the T-duality
among solutions with more than one vector/hyper multiplet, starting for
example with non-axionic black holes on the IIA side.
Just as the IIA black holes,
the IIB D-instanton solutions 
are then determined by a set of harmonic
functions plus the topological data
of the underlying Calabi-Yau space, like the intersection numbers.

Our paper is organized as follows. In the next section
we will first review the construction of the D-instanton solutions
of 10-dimensional type IIB superstring; we will consider
the dimensional reduction of $(0,4,4,4)$ IIA brane black hole brane
configurations together with their dual $(-1,3,3,3)$ IIB compactified
D-instanton configurations.
In section 3 we will discuss the effective $N=2$ supergravity
Lagrangian of $N=2$ vector and hyper multiplets focussing on
the c-map between the special K\"aher geometry of the vector fields
and the quaternionic geometry which is coming from the hyper multiplet
fields. Though this material is known in the literature we
have reviewed it with emphasis on the T-duality perspective
to make the presentation self-contained; 
the behaviour of the universal sector under
the c-map is investigated with particular care.
In section 4,
the main part of the paper, we will apply the T-duality, 
or c-map respectively,
to relate the IIA black hole solutions to the IIB D-instanton solutions.
Section 5 contains our conclusions.

\section{Ten-dimensional black holes and D-instantons}

\subsection{D-branes and D-instantons in ten dimensions}

The notion of a D-brane arises, when one considers open strings with 
Dirichlet boundary conditions along $10-p-1$ directions.
Then their endpoints are constrained to live on a 
$(p+1)$-dimensional submanifold, which is interpreted as the world volume
of a $p$-dimensional extended non-perturbative object, called
a D(irichlet)-$p$-brane, and which is thought of as a stringy soliton
or instanton. Such D-$p$-branes are the carriers
of R-R charges, which are predicted by string duality but
absent in type I or type II perturbation theory. To be
specific, the $(p+1)$-dimensional worldvolume can couple
to $(p+1)$-form gauge potential, implying that they are
charged under the corresponding $(p+2)$-form field strength.
Since one can choose Neumann or Dirichlet
boundary conditions along each space-time direction independently, one
expects that D-branes can exist for any values 
$-1 \leq p \leq 9$ of $p$ and  should exist if the corresponding
gauge field exists in the perturbative spectrum.

Within the low energy effective action, D-branes
can be identified with BPS-saturated, R-R charged 
$p$-brane solutions. Such $p$-brane solutions can
be characterized in terms of a function $H$,
which is harmonic with respect to the directions
transversal to the world volume. To describe a single
$p$-brane this function must be taken to depend only 
on the transversal radius $r=\sqrt{ {\bf x}_{t} \cdot
{\bf x}_{t}}$, i.e. $H=H(r)=c_{1} + \frac{c_{2}}{r^{7-p}}$.
Then the geometry in the string frame is 
\be0500
ds^{2}_{p} = \frac{1}{\sqrt{H}} (-dt^{2} + (dx^{1})^{2} 
+ \cdots + (dx^{p})^{2}) + \sqrt{H} ( (dx^{p+1})^{2} +
\ldots + (dx^{9})^{2} )
\eq
and the dilaton is
\be0510
e^{-2 \Phi} = H^{\frac{p-3}{2}} \,.
\eq
The gauge fields can likewise be expressed in terms of 
$H$ \hcite{BerDeR} 
\be0511
F = \left\{ \begin{array}{ll} d \frac{1}{H} 
\wedge dt \wedge dx^1\wedge \cdots \wedge
dx^p &
 , \ \mbox{for}\  p \leq 3 \\
 {^{\star}d} \frac{1}{H} \wedge dt \wedge dx^1\wedge \cdots \wedge dx^p &
 , \ \mbox{for}\  p \geq 3 
\end{array} \right. 
\eq 
and the field strengths fall of
as $\frac{1}{r^{8-p}}$ for $r \rightarrow \infty$, which
is the characteristic power law for a charged $p$-brane
in ten dimensions.

Since T-duality for open strings \cite{DaiLeiPol,AlvBarBor,BerDeR}
mutually exchanges Neumann and
Dirichlet boundary conditions, $p$-branes can be related
to $(p+1)$-branes by dualizing over a transversal direction  
and to $(p-1)$ branes by dualizing over a world volume 
direction. {From} (\href{0500}) and (\href{0510}) it is obvious
that T-duality acts appropriately on the metric 
and the dilaton. The action on the gauge fields has been 
worked out in \hcite{BerDeR}. Note that T-dualization changes  
the number of transversal directions, on which $H$ 
is allowed to depend. Thus, in order to go from a single
localized $p$-brane to a single, localized $(p \pm 1)$-brane
we have to restrict or relax the functional dependence of $H$, such
that it depends on the new transversal radius. This is always
understood when saying that a $p$-brane is T-dual to a
$(p \pm 1)$-brane.
We would also like to recall that in the case of type II 
strings T-duality exchanges type IIA with type IIB and vice versa 
\hcite{DaiLeiPol}.

Let us now recall the massless perturbative spectra and 
the known D-$p$-brane solutions of ten-dimensional 
type II string theories. The massless perturbative NS-NS spectrum
is the same for IIA and IIB. It consists of the metric
$G_{MN}$, the antisymmetric tensor $B_{MN}$ and the dilaton
$\Phi$.
The  R-R spectrum of IIA contains a 1-form $A_{M}$ and a
3-form $A_{MNP}$ whereas the massless R-R spectrum of IIB consists of
a 0-form $A$, a 2-form $A_{MN}$ and a 4-form
(with selfdual field strength) $A_{MNPQ}$. This  implies that
IIA (IIB) has only D-$p$-branes with even (odd) $p$.
The type II D-branes with $0 \leq p \leq 6$ correspond
to standard $p$-brane solitons of the IIA / IIB
effective supergravity actions. For $p=0,1,2$  they carry 
electric and for $p=6,5,4$ they carry magnetic charge
under the 1-, 2-, 3-form gauge potential,
respectively, whereas the selfdual 3-brane couples to the
selfdual 4-form. We refer to \hcite{DufKhuLu} for a review
and more references.

The $(-1)$-brane and the 7-brane have special
properties, that distinguish them from the other branes.
The associated gauge field is the 0-form $A$, which has
an axion like shift symmetry $A \rightarrow A +$ constant. 
Therefore it only enters physics via its derivatives $F_M = \p_{M} A$
which play the role of a field strength.
A more conventional gauge theory 
description arises when dualizing $A$ into a 8-form
$A^{(8)}$ with 9-form field strength $F^{(9)}=dA^{(8)}$ and
gauge invariance $A^{(8)} \rightarrow A^{(8)} + d \Lambda^{(7)}$
where $\Lambda^{(7)}$ is an arbitrary 7-form.
The $(-1)$-brane (7-brane) carries electric charge with
respect to the 0-form (8-form) and magnetic charge with
respect to the 8-form (0-form).

The D-7-brane solution \hcite{GibGrePer}
is the ten-dimensional analogue of the four-dimensional 
stringy cosmic string \hcite{GreShaVafYau} and therefore it 
is not asymptotically
flat. Moreover it plays an important role in the construction of 
F-theory \hcite{Vaf}.
On the other hand the $(-1)$-brane solution \hcite{GibGrePer}
is most naturally interpreted as an instanton, 
and has to be identified with the D-instanton of \hcite{Dinstanton}.
Thus one has to take time in (\href{0500}) to be imaginary.
We will consider this solution in more detail in the next 
subsection. 
Both the $(-1)$-brane and the 7-brane
are completely non-singular in the string frame.


Finally we would like to mention that 8-branes and
9-branes, for which no gauge potential is present
in the standard perturbative action have also been considered in the 
literature \hcite{Pol,BerDeR}.

\subsection{The ten-dimensional D-instanton}

Let us now recall the ten-dimensional D-instanton
solution of IIB supergravity \hcite{GibGrePer}.
We will focus on the crucial role of boundary terms
in the Euclidean domain. The discussion of other aspects
of the solution is more sketchy. For more details we 
refer the reader to \hcite{GibGrePer} and to the
discussion of the four-dimensional D-instanton solution
in section 4.1.

The action for the non-vanishing fields can be given
either in terms of the R-R zero-form $A$
\be0520
S = \int d^{10}x \sqrt{G} \left\{ 
R - \frac{1}{2} (\p \Phi)^{2} - \frac{1}{2} e^{2 \Phi} 
(\p A)^{2}  \right\} 
\eq
or in terms of the Hodge-$\star$-dual 9-form 
$F^{(9)}=e^{2 \Phi} \star d A$
\be0530
S' = \int d^{10}x \sqrt{G} \left\{ 
R - \frac{1}{2} (\p \Phi)^{2} - \frac{1}{2\cdot 9!} e^{-2 \Phi}
(F^{(9)})^{2} \right\}\,,
\eq
where $R$ is the curvature scalar, $G$ the absolute value of the
determinant of the metric and $\Phi$ is
the dilaton. The actions above refer to the Einstein frame with
signature $(-+\cdots +)$. When looking for instanton solutions
one has to Wick-rotate the theory to Euclidean signature. 
We take the Euclidean action to be positive. Then,
within our conventions, it is obtained from the Minkowski
action by replacing all fields by their Euclidean version and
putting an overall minus sign.
Note however that some subtleties arise, which are related to 
the fact that this Euclidean continuation
does not commute with $\star$-dualization.
This has the following effect: Dualizing the Euclidean
version of (\href{0530})
\be0540
S'_{Euc} = \int d^{10}x \sqrt{G} \left\{ 
-R + \frac{1}{2} (\p \Phi)^{2} + \frac{1}{2\cdot 9!} e^{-2 \Phi}
(F^{(9)})^{2} \right\}
\eq
yields
\be0550
S_{Euc} = \int d^{10}x \sqrt{G} \left\{ 
- R + \frac{1}{2} (\p \Phi)^{2} - \frac{1}{2} e^{2\Phi}
(\p A)^{2} \right\} + \oint e^{2 \Phi} A \wedge \star d A \,,
\eq
where the Euclidean 1-form field strength
$dA$ has been defined by $\star$-dualization
of the Euclidean 9-form $F^{(9)}$.
The Euclidean action (\href{0550}) differs from a naive
Wick rotation of (\href{0520}) in two respects as pointed out
in \hcite{GibGrePer} and in the appendix of \hcite{GreGut}: 
First 
the kinetic term of $A$ has a flipped sign, which
makes the sum of the scalar kinetic terms in the bulk part
of action (\href{0550}) indefinite. 
Second there is a boundary term, which is
crucial for finding the correct instanton action. Note
that the dual Euclidean 9-form action (\href{0540})
is completely standard.

The D-instanton solution can be derived from either
(\href{0540}) or (\href{0550}). In terms of the 0-form
$A$ it takes the following form \hcite{GibGrePer}:
First one requires that half of the Euclidean 
IIB supersymmetries remain unbroken in order to get
a BPS saturated configuration. This can be achieved by
relating $A$ to the dilaton $\Phi$ by
\be0560
d A = \pm e^{-\Phi} d \Phi \,.
\eq
With this ansatz the equations of motion are solved
by taking the Einstein metric to be flat and $e^{\Phi}$
to be harmonic. Thus a single D-instanton is described by
\be0570
e^{\Phi} = e^{\Phi_{\infty}} + \frac{c}{r^{8}} = H(r)\,.
\eq
$e^{\Phi}$ is singular at $r=0$, whereas the metric in the 
string frame
\be0580
ds^{2}_{-1} = e^{\Phi / 2} ((dt)^{2} + (dx^{1})^{2} + \ldots
+(dx^9)^2)
= \sqrt{H} (dr^{2} + r^{2} d\Omega_{9}^{2})
\eq
is asymptotically flat for $r \rightarrow 0$. In fact
it describes a finite neck wormhole connecting two 
asymptotically flat regions at $r\rightarrow 0, \infty$
\hcite{GibGrePer}.

Plugging this solution into (\href{0540}) or (\href{0550})
one obtains the finite instanton action
\be0590
S_{Inst} = - \oint \star d \Phi = \frac{|Q^{(-1)}|}{g} \,,
\eq
where the integration is over asymptotic nine-spheres
at $r=0$ and $r= \infty$, $Q^{(-1)}$ is the electric
charge of the solution with respect to $A$, (i.e. the Noether charge
associated to the 0-form gauge symmetry $A \rightarrow A+$ constant)
and $g=e^{\Phi(\infty)}$ is the string coupling at $\infty$.
(By explicit computation one finds that the boundary at $r=0$ does
not contribute to the action.)
The instanton action has a factor $\frac{1}{g}$ in front,
which is characteristic for its origin from the R-R sector.
Note that when using action (\href{0550}) the bulk part
vanishes and the whole instanton action comes from
the boundary term. This is the reason why the boundary
term should be kept despite of not contributing to the
equations of motion.

\subsection{The $(0,4,4,4)$ solution of IIA}

By T-duality the D-instanton solution of IIB can be
related to a (Euclideanized) 0-brane solution of IIA, i.e.
to a black hole. When later studying D-instanton like
solutions of $D=4$, $N=2$ supergravity and string theory
we will make use of the fact that extremal black holes
of these theories have been studied extensively in the last
two years.
Using T-duality on a general black hole (or even
stationary) solution we can then generate a variety of
instanton solutions. In order to get non-degenerate
black holes and instantons in four dimensions it is not
sufficient just to compactify ten dimensional 0-brane and 
$(-1)$-brane solutions, but one has to add further 4-branes and
3-branes. This will be recalled and explained in the next two
subsections.

Let us consider how to construct four-dimensional black
hole solutions out of ten-dimensional ones. This can be
done starting from (\href{0500}) for $p=0$ by toroidal
compactification. Then $H$ has to be a harmonic
function with respect to the remaining three transversal
spatial direction and thus behaves like $\frac{1}{r}$
for $r \rightarrow 0$. This implies that the asymptotic
sphere at $r=0$ has vanishing area, and one obtains a 
black hole with degenerate horizon and vanishing 
Bekenstein-Hawking entropy. To get a finite horizon one needs
to have more species of charges, leading to more harmonic
functions in the solution. In four dimensions one needs
precisely four charges to obtain a finite horizon.

When starting in ten dimensions such solutions can be 
constructed by considering a configuration of three
4-branes and one 0-brane with the following transversal
intersection pattern \hcite{SupHarmFunc}
\be0610
\begin{array}{c|cccccccccc}
 & t & y_1 & y_2 & y_3 & y_4 & y_5 & y_6 & x_1 & x_2 & x_3 \\ \hline
0 & \times&     &     &     &     &     &     &     &     &     \\  
4 & \times&     &     &\times   & \times  & \times  & \times  &     
  &     &     \\  
4 & \times&\times   & \times  &     &     & \times  & \times  &    
   &     &     \\  
4 & \times& \times  & \times  &\times   & \times  &     &     &     
  &     &     \\  
\end{array}
\eq
In this table world volume directions of the various
branes are marked by a $\times$. A BPS saturated configuration
solving the IIA equations of motion and having this
intersection pattern can be found and it depends on
four functions $H_{0},\ldots,H^{3}$, which are harmonic
with respect to the three overall transversal directions
${\bf x}= (x_{1},x_{2},x_{3})$ (and independent of the others).
The $D=10$ string frame metric is \hcite{SupHarmFunc} 
\be0620
d\,s^2 = \frac{-1}{\sqrt{H_0 H^1 H^2 H^3}} \, d \, t^2
+ \sqrt{H_0 H^1 H^2 H^3} \,d \, {\bf x}^2 + 
\eq
\[
+ \sqrt{ \frac{H_0 H^1}{H^2 H^3} } \left( d\,y_1^2 + d\,y_2^2 \right)
+ \sqrt{ \frac{H_0 H^2}{H^1 H^3} } \left( d\,y_3^2 + d\,y_4^2 \right)
+ \sqrt{ \frac{H_0 H^3}{H^1 H^2} } \left( d\,y_5^2 + d\,y_6^2 \right)
\]
and the ten-dimensional dilaton is
\be0630
e^{-2 \Phi} = \sqrt{\frac{H^1 H^2 H^3}{H_0^3}} \,.
\eq

The gauge field strengths associated with the (electric) 0-brane
and with the first of the (magnetic) 4-branes in (\href{0610})
are
\be0640
F^{(2)} = d\,\frac{1}{H_{0}} \wedge d\,t  \;\;\;
\mbox{and} \;\;\;
F^{(4)} = * d \frac{1}{H^{1}} \wedge d\,y_{3} \wedge d\,y_{4}
\wedge d\,y_{5} \wedge d\,y_{6} \wedge d\,t  \,.
\eq 
The expressions for the remaining 4-branes are obtained by
obvious replacements of coordinates.

Compactifying to $D=4$ one obtains a black hole with
metric 
\be0650
d\,s_4^2 = \frac{-1}{\sqrt{H_0 H^1 H^2 H^3}} \,d\,t^2
+ \sqrt{H_0 H^1 H^2 H^3} \, d\,{\bf x}^2
\eq
and constant four-dimensional dilaton $\varphi$
\be0660
e^{-2 \varphi} = e^{-2 \Phi} \sqrt{G^{(int)}} = 1 \,.
\eq
Equation (\href{0660}) implies that string and Einstein
frame coincide. 
The black hole (\href{0650})
is extremal, carries one electric charge $q_0$
and three magnetic 
charges $p^1$, $p^2$, $p^3$
and has a finite horizon. The geometry is 
similar to the extreme Reissner-Nordstrom black hole.
The harmonic functions take the form
\be0665
H_0 = h_0 + \frac{q_0}{r},\;\;\;
H^A = h^A + \frac{p^A}{r},\;\;\;
r = \sqrt{x^2 + y^2 + z^2}\,.
\eq
When taking all the harmonic functions to be equal,
$H_0 =H^1 =H^2 =H^3$, then (\href{0650}) precisely describes
the (outer part of the) extreme Reissner Nordstrom geometry with the
horizon located at $r=0$. To see that the horizon is finite one
looks at the behaviour of (\href{0650}) for $r \rightarrow 0$, using
that the asymptotic behaviour of the harmonic functions:
\be0655
\sqrt{H_0 H^1 H^2 H^3} \, d\,{\bf x}^2
\longrightarrow 
\frac{\sqrt{q_0 p^1 p^2 p^3}}{r^2} \left( 
(dr)^2  +  r^2 d \Omega_2^2 \right)
\eq
This shows that one has to start with four charged objects in
$D=10$ in order to have a black hole with finite horizon in
$D=4$. The geometry of the six internal dimensions
enters the four-dimensional effective action through the 
three moduli scalars 
$T_{1} = \sqrt{ \frac{H_0 H^1}{H^2 H^3} }$,
$T_{2}=\sqrt{ \frac{H_0 H^2}{H^1 H^3} }$ and 
$T_{3} = \sqrt{ \frac{H_0 H^3}{H^1 H^2} }$, which parametrize
the size of three (pairwise orthogonal) internal 2-tori.
For a generic choice of the four harmonic functions
the moduli are space dependent, i.e. the geometry of the
internal space varies when one moves around in four-dimensional
space-time.

\subsection{The $(-1,3,3,3)$ solution of IIB}

One can now rotate the $(0,4,4,4)$ solution of IIA
to Euclidean time and then T-dualize it over the time
direction. The result is a configuration with one 
$(-1)$-brane and three 3-branes with the following 
intersection pattern \hcite{SupHarmFunc}:
\be0670
\begin{array}{c|cccccccccc}
  & t & y_1 & y_2 & y_3 & y_4 & y_5 & y_6 & x_1 & x_2 & x_3 \\ \hline
-1&   &     &     &     &     &     &     &     &     &     \\  
3 &   &     &     &\times   & \times  & \times  & \times  &     &     
&     \\  
3 &   &\times   & \times  &     &     & \times  & \times  &     &     
&     \\  
3 &   & \times  & \times  &\times   & \times  &     &     &     &     
&     \\  
\end{array}
\eq 
The metric in the ten-dimensional string frame is
\be0680
d\,s^2 = \sqrt{H_0 H^1 H^2 H^3} d\,t^2
+ \sqrt{H_0 H^1 H^2 H^3} d\,{\bf x}^2 
\eq
\[
+ \sqrt{ \frac{H_0 H^1}{H^2 H^3} } \left( d\,y_1^2 + d\,y_2^2 \right)
+ \sqrt{ \frac{H_0 H^2}{H^1 H^3} } \left( d\,y_3^2 + d\,y_4^2 \right)
+ \sqrt{ \frac{H_0 H^3}{H^1 H^2} } \left( d\,y_5^2 + d\,y_6^2 \right)
\]
and the dilaton is
\be0690
e^{-2 \Phi} = \frac{1}{H_{0}^{2}} \,.
\eq
Note that all world volume directions are spacelike
so that the whole configuration is instanton-like.
The 0-form gauge potential $A$ is related
to the dilaton by the instanton ansatz (\href{0560})
which implies
\be0695
A = \mp H_0^{-1} + \mbox{constant}\,.
\eq
The selfdual 5-form field strength associated
with the first 3-brane in (\href{0670}) is
\be0700
F^{(5)} =  d\, \frac{1}{H^{1}} \wedge d\,y_{3} \wedge d\,y_{4}
\wedge d\,y_{5} \wedge d\,y_{6} + \star
(d\, \frac{1}{H^{1}} \wedge d\,y_{3} \wedge d\,y_{4}
\wedge d\,y_{5} \wedge d\,y_{6})
\eq
and similar expressions hold for the other 3-branes.
 
When compactifying to four dimensions we have to 
introduce the four-dimensional dilaton $\varphi$:
\be0710
e^{-2 \varphi} = e^{-2 \Phi} \sqrt{ G^{(int)} } =
(H_{0} H^{1} H^{2} H^{3})^{-1/2} \,.
\eq
The four-dimensional string frame and Einstein frame metrics are
\be0720
ds^{2} = e^{2 \varphi} ds^{2}_{E} = \sqrt{H_{0} H^{1} H^{2} H^{3}}
((dt)^{2} + (d {\bf x})^{2}) \,.
\eq
As we explained earlier the harmonic functions
$H_{0}, H^A$ should depend on the overall transversal radius in
order to describe a localized D-instanton. For our
four-dimensional case this means that 
they are functions of $r=\sqrt{ t^2 + {\bf x}^2 }$:
\be0721
H_0 = h_0 + \frac{q_0}{r^2},\;\;\;
H^A = h^A + \frac{p^A}{r^2} \,.
\eq
The simplest
solution is obtained by taking all four harmonic functions
(and therefore all the charges)
to be equal, implying that $e^{\varphi}$ harmonic:
\be0730
e^{\varphi} = H_{0} = H^{1} = H^2 = H^3 = e^{\varphi(\infty)}
+ \frac{q_0}{r^{2}}; \;\;\; \;\;\;
A= \mp \frac{1}{e^{\varphi_{\infty}} 
+ \frac{q_0}{r^{2}}} + \mbox{constant}.
\eq
This is the four-dimensional analogue of the ten-dimensional
finite neck wormhole discussed in section 2.2.
Note that one needs 4 harmonic functions in (\href{0720}) in order
to stabilize the neck of the wormhole. This is the same
condition that one has for black holes in order to get a
finite horizon. As in the black hole case the internal
geometry enters the four-dimensional theory through
the three non-trivial moduli fields
$T_{1} = \sqrt{ \frac{H_0 H^1}{H^2 H^3} }$,
$T_{2}=\sqrt{ \frac{H_0 H^2}{H^1 H^3} }$ and 
$T_{3} = \sqrt{ \frac{H_0 H^3}{H^1 H^2} }$. 

In the following we will generalize this simple setup in two
ways. First, the black hole solutions obtained
here by compactification from $D=10$ are not generic
$D=4$ solutions. In four dimensions 2-form gauge fields
are dual to 2-form gauge fields and therefore black hole 
solutions are in general dyonic. Moreover one can vary 
the values of various moduli scalar fields at infinity and
deform the solution.
This way one can for example turn on non--trivial
$\theta$-angles, leading to axionic black holes which 
are not easily described from
the ten-dimensional point of view. In this case it is appropriate to
start with the general $D=4$ action and solve the
corresponding equations  of motion. Secondly we have 
used toroidal compactification to go from ten to four 
dimensions. This procedure gives us solutions of $N=8$
supergravity in ten dimensions. We would, however, like
to have a smaller number of supersymmetries. In particular
we are interested in $N=2$,  
which corresponds to compactification on a 
Calabi-Yau threefold. 

Both kinds of generalization 
make it necessary to recall certain facts about
$D=4$, $N=2$ supergravity, which will be done in the
next section before returning to explicit solutions in the
following one.

\section{$N=2$, $D=4$ type II string theory on Calabi-Yau manifolds}
\setcounter{equation}{0}

\subsection{Ungauged $N=2$ supergravity: action, equations of motion
and special geometry}

The action of the effective $N=2$ supergravity theory that we want
to study in four dimensions will contain the following kinds
of $N=2$ multiplets: First there is the gravity
multiplet 
\be0002
(e^m_{\mu}, {\cal A}_{\mu}, \psi_{\mu}^{i})\;\;\;
m,\mu=0,\ldots,3,\;\;\;i=1,2\,,
\eq  
which contains the graviton, 
the graviphoton (with field strength $T_{\mu \nu})$ 
and two gravitini. 
Second there are $N_V$ vector multiplets 
\be0004
({\cal A}^A_{\mu}, z^A, \lambda^A_{i}), \;\;\;
A=1,\ldots,N_V\,,i=1,2\,,
\eq
which consist of a vector (with field strength ${\cal F}^A_{\mu \nu}$),
a complex scalar and two gauginos, which are Weyl spinors.
Third there are $N_H$ hyper multiplets
\be0006
(q^{u}, \xi_{\alpha}),
\;\;\; u=1,\ldots,4 N_H,\;\;\;\alpha=1,\ldots,2N_H
\eq
with 4 real scalars and two Weyl spinors.

The structure of $N=2$ supergravity coupled to vector and
hyper multiplets is governed by special geometry
(for a review see \cite{WVV,ABC,CRTP}).
This means that the action is completely specified 
in terms of three geometrical data: 
\begin{enumerate}
\item
The special K\"ahler manifold $K_{N_V}$ parametrized by the
vector multiplet scalars $z^A$. It
describes the coupling and self-interactions of $N_V$ vector multiplets
to $N=2$ supergravity \cite{SpecK1,special_K,CRTP}. 
\item
The quaternionic manifold $Q_{N_H}$ parametrized by the hyper multiplet
scalars $q^u$.
It describes the coupling and self-interactions of 
$N_H$ hyper multiplets
to $N=2$ supergravity \cite{ba/wi,deWLauPro,special_H,WVV}. 
\item
The choice of the gauge group ${\cal G}$.
\end{enumerate}

For type II Calabi-Yau compactifications one gets at generic points
in the moduli space what is called 
'ungauged supergravity', i.e. all the vector multiplets
are abelian and all hyper multiplets are neutral.
The gauge group is ${\cal G} = U(1)^{N_V+1}$, where the additional 
$U(1)$ is due to the graviphoton. One important issue
of matter coupled $N=2$ supergravity is, reflecting the
points (1) and (2), that the moduli space is a product space
\begin{eqnarray}
{\cal M} &=& K_{N_V} \ \otimes \  Q_{N_H}.
\end{eqnarray}  
Vector and hyper multiplet can only couple through gravity
and through gauge couplings. The later ones are absent in the ungauged case.

As already mentioned above
$N=2$ supergravity imposes further geometric constraints on the scalar 
manifold
besides factorization, which define what is called special
geometry. The vector multiplet moduli space must be a special 
K\"ahler manifold. This is a K\"ahler-Hodge
manifold
with the additional properties that (i) the K\"ahler potential can
be obtained from a holomorphic section $(X^I(z), F_I(z))$ of a
symplectic vector bundle over the manifold by
$K(z,\ov{z})=-\log(i(\ov{X^I}(\ov{z}) F_I(z) - X^I(z) \ov{F_I}(\ov{z})))$
and (ii) that this section satisfies
$X^I(z) \p_A F_I(z) = F_I(z) \p_A X^I(z)$ \hcite{CRTP}.  All physical
quantities
of the vector multiplet sector and in particular the couplings
appearing in the action can be obtained from the section.
In case that the matrix $(N_V+1) \times (N_V+1)$ matrix 
$M_{IJ}$, which is defined by 
$M_{AI}= D_A X^I$ and  $M_{0I} = X^I$ is invertible, 
one can obtain the section
(and thus everything) from a prepotential $F(X^I)$ by
$F_I = \p F /\p X^I$ \hcite{CDFP,CRTP}. 
(Here $D_A$ denotes the space-time and
K\"ahler covariant derivative.)
The prepotential is holomorphic and homogenous of
degree 2 in $X^I$ \hcite{SpecK1}. One can always go to a section which comes
{from} a prepotential by a symplectic transformation, and for the
purposes of this paper such symplectic transformations will not
change the physics of the model (see however: \cite{CDFP,FGP}).

The hyper multiplet moduli space $Q_{N_H}$ is likewise subject to 
geometric 
restrictions. It was shown in \cite{ba/wi} that the following
conditions must hold:
(i) the holonomy group must be contained in $SU(2) \otimes Sp(2 N_H)$,
(ii) the $SU(2)$ part of the curvature must be non--trivial. Such manifolds
are called quaternionic because they simultanously admit 
three complex structures
$J_i$, $i=1,2,3$, such that the metric is hermitean with respect
to all three of them and they satisfy the quaternionic
algebra
\be0008
J_{i} J_{j} = - \delta_{ij} {\bf 1} + \epsilon_{ijk} J_k \,.
\eq
The associated hyper K\"ahler form, i.e. the $SU(2)$ triplet of
K\"ahler forms corresponding to the three complex structures
is proportional to the $SU(2)$ curvature. In the next section
we will show in a simple example how the hyper K\"ahler form and
the three complex structure can be obtained from the $SU(2)$ connection.
It might be useful to remark that some authors, for example \hcite{ABC},
prefer to define quaternionic manifolds
in terms of the existence of three complex
structures and their properties.  
Then the structure of the holonomy  
can be deduced from the definition. We have followed here reference
\hcite{ba/wi} which defines 
a quaternionic manifold by the constraint on the holonomy group.
Then the existence of three complex structures with certain properties
is a consequence.

Before formulating the final condition we have to recall that 
quaternionic spaces are Einstein spaces, i.e. they have 
a constant curvature scalar $R$. (The case of one quaternionic 
dimension is somewhat special in that the holonomy constraint is trivial and
one has to add this condition by hand \cite{ba/wi}.)
Quaternionic manifolds describing hyper multiplet couplings to
$N=2$ supergravity have to obey an addtional constraint: (iii)
the curvature scalar must be negative and is fixed by the number 
of hyper multiplets (i.e. the dimension):
\bea0009
{\cal R} [Q_{N_H}] &=& -8 \ (N_H^2 + 2 N_H).
\eqa
We will in the following take this extra condition as part of the 
definition of quaternionic.

Let us now recall the bosonic part of the action of ungauged
$N=2$ supergravity with $N_V$ vector and $N_H$ hyper multiplets 
\hcite{ABC}
\be0010
S = \int d^4 x \, \sqrt{G} \, \left\{
R - 2 g_{A \ov{B}} \p_{\mu} z^A \p^{\mu} \ov{z^B}-
\frac{1}{4} \left( \Im {\cal N}_{IJ} F^I_{\mu \nu}  F^{J|\mu \nu}
+ \Re {\cal N}_{IJ}
F^I_{\mu \nu} {^{\star}F^{J|\mu \nu}} \right)
- \wt{g}_{uv} \p_{\mu} q^u \p^{\mu} q^v \right\}\,.
\eq
Here $G_{\mu \nu}$ is the space-time metric
in the Einstein frame, $G = |\det G_{\mu \nu}|$
and $R$ the corresponding curvature scalar.
The $N_V$ complex scalars $z^A$ coming from vector multiplets
parametrize  a special K\"ahler manifold $K_{N_V}$
with metric $g_{A \ov{B}}(z, \ov{z})$,
whereas the $4N_H$  
 real scalars $q^u$ parametrize the quaternionic manifold $Q_{N_H}$
with metric $\wt{g}_{uv}(q)$.
$F^I_{\mu \nu}$ are the field strength of the $N_V+1$ vectors in a 
symplectic basis, ${}^{\star}F^I_{\mu \nu}$ the (Hodge-) dual
field strength, ${\cal N}_{IJ}(z,\ov{z})$ is the gauge kinetic matrix.
Its imaginary part $\Im {\cal N}_{IJ}$ is positive definite and 
generalized the gauge couplings $\frac{1}{g^2_{YM}}$ of ordinary Yang-Mills
theory, wheras the real part $\Re{\cal N}_{IJ}$ is not restricted
and acts as a generalized (field dependent) $\Theta$ angle.

The symplectic field strength $F^I_{\mu \nu}$ are related
to the field strength $T_{\mu\nu},{\cal F}^A_{\mu\nu}$ of the
graviphoton and of the $N_V$ vectors sitting in vector multiplets 
as follows:
The (antiselfdual part of the) graviphoton field strength is given
in terms of the section $(X^I, F_I)$
by the symplectic invariant combination
$T^-_{\mu\nu} = F_J F^{- J| \mu \nu} - X^I G^{-}_{I|\mu \nu}$,
where $G^{-}_{I|\mu \nu} = \ov{\cal N}_{IJ} F^{-|J}_{\mu \nu}$.
On the other hand the $N_V$ field strength of vector multiplets are
${\cal  F}^{-|A}_{\mu \nu} = g^{A\ov{B}}( D_{\ov{B}} \ov{F_J} F^{-|J}_{\mu \nu}
- D_{\ov{B}} \ov{X^I} G^{-}_{I|\mu \nu})$. 
The vector multiplet couplings $g_{A\ov{B}}$ and ${\cal N}_{IJ}$
can be obtained from the holomorphic section $(X^I(z), F_I(z))$
this way:
The K\"ahler metric is derived from the K\"ahler potential by
$g_{A\ov{B}} = \p_A \p_B K$ with 
$K = -\log(i( \ov{X^I}(\ov{z}) F_I(z) - X^I(z) \ov{F_I} (\ov{z})))$,
whereas the gauge kinetic matrix ${\cal N}_{IJ}$ can be computed as 
follows: First one defines $f_A^I = D_A X^I$ and 
$h_{AI} = D_A F_I$, where $D_A$ is the K\"ahler covariant
derivative (see the references quoted above for the details).
Then one defines two $(N_V + 1) \times (N_V + 1)$ matrices $f=(f_J^I)$,
$h=(h_{JI})$ by introducing $f_0^I = \ov{X}^I$ and $h_{0I} = \ov{F}_I$.
Finally the gauge kinetic matrix is given by 
${\cal N}_{IJ} = \ov{h}^{T}_{IK} \ov{f}^{-1}_{KJ}$. In case that
we choose the section such that a prepotential exists, ${\cal N}_{IJ}$
can be computed by the more familiar formula
\be0018
{\cal N}_{IJ} = \ov{F}_{IJ} + 2 i \frac{ \Im F_{IK} \Im F_{JL} X^K X^L}
{ \Im F_{KL} X^K X^L }
\eq
where $F_{IJ} = \frac{ \p^2 F}{ \p X^I \p X^J}$.

For later use we will next list the equations of motion for the case
of vanishing fermions.
The gravitational one is:
\be0020
R_{\mu \nu} = 2 g_{A \ov{B}} \p_{(\mu} z^A \p_{\nu)} \ov{z^B}
+\wt{g}_{uv} \p_{(\mu} q^u \p_{\nu)} q^v 
+ \frac{1}{2} ( \Im {\cal N}_{IJ} F^I_{\mu \rho} F^{J|\rho}_{\nu}
+ \Re {\cal N}_{IJ} F^I_{\mu \rho}  {}^{\star}F^{J|\rho}_{\nu} )
\eq
\[
- \frac{1}{8} G_{\mu \nu} 
( \Im {\cal N}_{IJ} F^I_{\rho \sigma } F^{J|\rho \sigma}
+ \Re {\cal N}_{IJ} F^I_{\rho \sigma}  {}^{\star}F^{J|\rho \sigma } )
\]

The equation for the scalars
$q^u$,
\be0025
\frac{2}{\sqrt{G}} \p_{\mu} \left( \sqrt{G} \wt{g}_{uw} \p^{\mu} q^u
\right) - \p_w \wt{g}_{uv} \p_{\mu} q^u \p^{\mu} q^v = 0 \,,
\eq
can be rewritten using the definition of Christoffel symbols 
in a suggestive form:
\be0030
\Box_G q^w + \wt{\Gamma}_{uv}^w \p_{\mu} q^u \p^{\mu} q^v = 0 \,.
\eq
This looks like a generalized geodesic equation, in which 
4-dimensional curved space-time is mapped by the scalar fields $q^u$
into moduli space. (Note that $\Box_G$ is the Laplace operator
in space-time, whereas $\Gamma_{uv}^w$ are the Christoffel symbols in
moduli space.)

The equations for the 
scalars $z^A$ can be brought to a similar form,
\be0040
\Box_G z^D + \Gamma_{AB}^D \p_{\mu} z^A \p^{\mu} z^B = 
\frac{1}{8} g^{D \ov{E}} \left(
\p_{\ov{E}} \Im {\cal N}_{IJ}
F^I_{\mu \nu} F^{J|\mu \nu} +\p_{\ov{E}}  \Re {\cal N}_{IJ}
F^I_{\mu \nu} {}^{\star} F^{J|\mu \nu} \right)\,,
\eq
where this time the equation is inhomogenous by 
contributions of the gauge fields. This inhomogenity could
be interpreted as a potential that modifies the geodesic motion of
the $z^A$.
Note that the r.h.s. is not only absent if there are no gauge fields,
$F^I_{\mu \nu} = 0$, but also for holomorphic ${\cal N}$,
$\p_{\ov{E}} {\cal N} =0$. The gauge kinetic matrix ${\cal N}$ is
holomorphic in $z^A$ if and only if the prepotential is quadratic .
This is the case of minimal coupling which does not occur in
type II compactifications \hcite{cec}.

Finally we have the equations of motions for the gauge fields,
\be0050
\nabla_{\mu} ( \Im {\cal N}_{IJ} F^{J|\mu \nu} + 
\Re {\cal N}_{IJ} {}^{\star} F^{J|\mu \nu} )= 
0\,, 
\eq
which have to be supplemented by the Bianchi identities,
\be0060
\nabla_{\mu} {}^{\star} F^{I|\mu \nu} = 0\,.
\eq


\subsection{Type II compactifications, T-duality and the c-map}


In the last section we dealt with generic properties of ungauged $N=2$
supergravity. Let us now consider the specific case of 
compactifications of IIA and IIB superstrings
on Calabi-Yau threefolds. 
(In fact all this can be formulated for general type II compactifications
with general internal conformal field theories that are only
constrained to give $N=2$ space time supersymmetry in four 
dimensions.)
Let us
recall the spectra of these theories 
and their relation to Calabi-Yau geometry (see for example 
\hcite{FerThei}).
Since we know that we have $N=2$ supersymmetry in $D=4$ we
only need to consider the bosonic spectrum.

Consider first the IIA theory compactified on a Calabi-Yau 
threefold with Hodge numbers $h_{1,1}$ and $h_{2,1}$,
which count the numbers of independent
deformations of the K\"ahler and complex structure, respectively.
We split the indices into real space-time indices $\mu,\nu=0,\ldots,3$
and complex internal indices $i,j,\ldots = 1,2,3$.
The resulting $D=4$ bosonic spectrum is as follows: 
{From} the ten-dimensional metric $G_{MN}$ and torsion
$B_{MN}$ we get the four-dimensional graviton
$G_{\mu \nu}$, the four-dimensional axion $\wt{\phi} \sim B_{\mu \nu}$
$h_{2,1}$ complex scalars $G^K_{ij}$, $K=1,\ldots,h_{2,1}$ 
describing deformations of the complex structure of the internal 
manifold
and $h_{1,1}$ complex scalars $z^A \sim (G^A_{i\ov{j}} + i B^A_{i\ov{j}})$,
$A=1,\ldots,h_{1,1}$,
related to deformations of the complexified K\"ahler structure
of the internal manifold. In addition one gets the four-dimensional
dilaton $\varphi$ from the ten-dimensional one.
Next we look at the R-R sector: The 1-form $A_M$ gives a vector $A_{\mu}$
and from the 3-form $A_{MNP}$ we get  
$h_{1,1}$ vectors $A^A_{\mu i \ov{j}}$ together with
$h_{2,1} + 1$ complex scalars $A^K_{ij \ov{k}}$ and
$A_{ijk}$. 
These fields combine with fermions into the $N=2$
gravity multiplet, $N^{(A)}_{V}= h_{1,1}$ vector multiplets and
$N^{(A)}_{H} = h_{2,1} + 1$ hyper multiplets. There is one special
hyper multiplet that does not contain any moduli of the Calabi-Yau but
the four-dimensional coupling (dilaton) $\varphi$ together with the
axion $\wt{\phi}$ and two real R-R
scalars $\zeta^0$, $\wt{\zeta}_0$ corresponding to the complex scalar 
$A_{ijk}$. It is called the universal hyper multiplet. 

Let us now consider IIB supergravity on the same Calabi-Yau
threefold. In $D=10$ we have the same NS-NS fields as for IIA.
In the R-R sector we get the following: The 0-form $A$ gives
a scalar $\zeta^0$. {From} the 2-form $A_{MN}$ we get $h_{1,1}+1$
scalars $\zeta^A \sim A_{i\ov{j}}^{A}$ and 
$\wt{\zeta}_0 \sim A_{\mu \nu}$
and from the selfdual
4-form $A_{MNPQ}$ we get $h_{2,1}+1$ vectors
$A_{\mu ijk}^K$, $A_{\mu ij \ov{k}}$ and $h_{1,1}$ scalars
$\wt{\zeta}_A \sim A_{\mu \nu i \ov{j}}^A$. 
These fields combine with fermions into 
the  gravity multiplet,
$N^{(B)}_V = h_{2,1}$ vector multiplets and 
$N^{(B)}_H = h_{1,1} + 1$ hyper multiplets.
The singled out universal hyper multiplet this time contains
besides the dilaton $\varphi$ and the axion $\wt{\phi}$ 
and the R-R scalars $\zeta^0$ and $\wt{\zeta}_0$.

Comparing the two spectra one realizes that the numbers of 
vector multiplets and hyper multiplets are (almost) exchanged:
\be0070
N^{(A)}_V =  h_{1,1} = N^{(B)}_H - 1,\;\;\;
N^{(A)}_H - 1 = h_{2,1} = N^{(B)}_{V} \,.
\eq
Since the ten-dimensional IIA and IIB theory are related by 
T-duality and since we compactified both IIA and IIB on the 
same space one might expect that this can be explained 
by four-dimensional T-duality. It has been shown some time
ago that this is indeed the case \hcite{crem,cec,ferr}. 
As a consequence, there exists
a mapping, called the c-map  which relates
the two moduli spaces:
\be0080
c: K^{(A)}_{h_{1,1}} \times Q^{(A)}_{h_{2,1}+1}
\longleftrightarrow
Q^{(B)}_{h_{1,1}+1} \times K^{(B)}_{h_{2,1}}
\eq

Let us sketch briefly how the T-duality transformation is 
performed and
how the c-map is obtained. We will follow \hcite{ferr}. For details
the reader can consult either this paper or go through the almost
identical procedure used in section 4.3 for T-dualizing a 
stationary solution into an instanton.

Consider first the gravity and vector multiplet 
sector of (\href{0010}) 
\be0082
S_{IIA}^{G/V} = \int d^4 x \sqrt{ G^{(IIA)} } \left\{
R - 2 g_{A \ov{B}} \p_{\mu} z^A \p^{\mu} \ov{z^B}-
\frac{1}{4} \left( \Im {\cal N}_{IJ} F^I_{\mu \nu}  F^{J|\mu \nu}
+ \Re {\cal N}_{IJ}
F^I_{\mu \nu} {^{\star}F^{J|\mu \nu}} \right) \right\}
\eq
For definiteness we have called this theory the IIA theory and we will 
call its dual the IIB theory. Note however that there is no fundamental
distinction between IIA and IIB in $D=4$ since by mirror
symmetry IIA on a Calabi--Yau is the same theory as IIB on 
the mirror manifold.

In $D=4$ we start
with the metric $G^{IIA}_{\mu \nu}$, $N_V+1$ physical vectors
$A^I_{\mu}$ 
(including the graviphoton)
and $N_V$ physical complex scalars $z^A$. 
Now compactify this action to $D=3$ along an isometry direction
and decompose the four-dimensional fields into three-dimensional ones.
{From} the metric we get a Kaluza-Klein scalar $\phi$ and a
Kaluza-Klein vector $\omega_{m}$ $m=1,2,3$. (The remaining three-dimensional
metric is non--dynamical, because there are no gravitons in three
dimensions.) Likewise every four-dimensional
vector gives a scalar $\zeta^I \sim A_0^I$ and a vector $A_m^I$.
And the scalars $z^A$ remain of course scalars. 
Now in three dimensions a vector can be dualized into a scalar.
Replacing the Kaluza Klein vector $\omega_m$ and the vectors
$A_m^I$ by scalars $\wt{\phi}$, $\wt{\zeta}_I$ we end up with 
a total of $4(N_V + 1)$ real scalars, which can be identified
with the $4(N_V+1)$ scalars that one gets by compactifying the hyper multiplet
scalars of the dual IIB theory on a circle of inverse radius.
After decompactification one gets the hyper multiplet part of the
dual action
\be0084
S_{IIB}^{H} = \int d^4 x \sqrt{G^{(IIB)}} \left\{
- 2 g_{A \ov{B}} \p_{\mu} z^A \p^{\mu} \ov{z}^{\ov{B}}
- \frac{1}{2 \phi^{2}} (\p \phi)^{2} - \frac{1}{2 \phi^{2}}
( \p \wt{\phi} +  \zeta^{I} \p \wt{\zeta}_{I}
- \p \zeta^{I} \wt{\zeta}_{I})^{2}
\right.
\eq
\[
\left.
- \frac{1}{\phi} \p \zeta^{I} \Im {\cal N}_{IJ} \p \zeta^{J} 
- \frac{1}{\phi} ( \p \wt{\zeta}_{I} + \Re {\cal N}_{IK}
\p \zeta^{K}) \Im {\cal N}^{IJ} 
(\p \wt{\zeta}_{J} + \Re {\cal N}_{JL} \zeta^{L})  \right\} \,,
\]
where ${\cal N}^{IJ}$ is the inverse of ${\cal N}_{IJ}$.
As shown in \hcite{ferr} this $\sigma$-model is indeed quaternionic.
We will come back to this point later. 

Thus one can map every special K\"ahler manifold
$K_{N_V^{(A)}}$ of complex dimension $N_V^{(A)}$ to a quaternionic manifold
$Q_{N_V^{(A)}+1}$ of quaternionic dimension $N_V^{(A)}+1 = N_H^{(B)}$:
\be0090
 s_{N_V^{(A)}}:   \hspace{1cm}    K_{N_V^{(A)}} \, 
 \rightarrow  \, Q_{N_V^{(A)} + 1}  
\eq
This is called the s-map. 
Note that one quaternionic dimension comes from the 4 bosonic degrees of 
freedom of the gravity multiplet. Quaternionic manifolds that can
be obtained this way are called special quaternionic manifolds.

On the other hand T-duality requires that every 
hyper multiplet moduli space arising from a type II compactification
must be special quaternionic, i.e. the corresponding  
scalar $\sigma$-model can be written as in (\href{0084}).
Reversing the above procedure one can map every such 
quaternionic manifold
of quaternionic dimension $N_H^{(A)}$ to a corresponding
special K\"ahler manifold of complex dimension $N_H^{(A)} -1$:
\be0100
s_{N_H^{(A)}}^{-1} : \hspace{1cm}    Q_{N_H^{(A)}} \, 
 \rightarrow  \, K_{N_V^{(A)} - 1}  \,.
\eq
This time $2 N_H^{(A)} + 4$ real scalars get mapped to the graviton and
to the $N_H^{(A)}=N_V^{(B)} + 1$ vector fields of the dual theory.
In particular the dynamical degrees of freedom of the IIB metric
in (\href{0084}) arise from two scalars in $S_{IIA}^{H}$ which
is dualized into $S_{IIB}^{G/V}$.

The full c-map is given by combining both sectors:
\be0110
c = s_{N_V^{(A)}} \times s_{N_H^{(A)}}^{-1}: \hspace{1cm}
K^{(A)}_{N_V} \times Q^{(A)}_{N_H}
\longrightarrow
Q^{(B)}_{N_V+1} \times K^{(B)}_{N_H -1}
\eq


Let us consider the s-map and the $\sigma$-model \href{0084} in some more
detail. Following \hcite{ferr} the action \href{0084} can 
rewritten in a way that makes the geometry more transparent and
allows one to show that the target space is quaternionic and
that certain subspaces are K\"ahler. This will be used later when
we dualize solutions of the equations of motion.

So we combine the $2N_V^{(A)}+4$ real scalars
$\phi, \wt{\phi}, \zeta^I, \wt{\zeta}_I$
into $N_V^{(A)} + 2$ complex ones \hcite{ferr}:
\bea0120
S' &=& \phi - i \zeta^I {\cal N}_{IJ} \zeta^J
+ i \wt{\phi} - i \zeta^I \wt{\zeta}_I \nonumber\\
C_I &=& - \Im {\cal N}_{IJ} \zeta^J + i ( \wt{\zeta}_I + \Re {\cal N}_{IJ}
\zeta^J) \\
\nonumber
\eqa
Rewriting the action (\href{0084}) in terms of the $2(N_V^{(A)} +1)$
complex scalars $S', C_I, z^A$ results in
\bea0130
S^{H}_{IIB} &=& \int d^4 x \sqrt{ G^{IIB}} \left\{
 -2 g_{A\ov{B}} \p z^A \p \ov{z^B} \right. \nonumber \\
 &- & 2 \frac{ | \p S' - (C+\ov{C})_I \Im {\cal N}^{IJ} \p C
- \frac{i}{4} (C+\ov{C})_I  \Im {\cal N}^{IJ} \p {\cal N}_{JK}
\Im {\cal N}^{KL} (C+\ov{C})_L |^2 }
{(S'+\ov{S'} - \frac{1}{2} (C+\ov{C})_I  \Im {\cal N}^{IJ}
(C+\ov{C})_J )^2} \nonumber \\
&-& \left. 2 \frac{ ( \p C_I + \frac{i}{2} \p {\cal N}_{IK} \Im {\cal N}^{KL}
(C+\ov{C})_L) \Im {\cal N}^{IJ}
( \p \ov{ C_J} - \frac{i}{2} \p \ov{\cal N}_{JM} \Im {\cal N}^{MN}
(C+\ov{C})_N)}
{S'+\ov{S'} - \frac{1}{2} (C+\ov{C})_I  \Im {\cal N}^{IJ}
(C+\ov{C})_J}  \right\} \\
\nonumber
\eqa

Starting from this expression one can define a vielbein
in terms of $S', C_I, z^A$ and show that the target space 
is quaternionic by computing the corresponding curvature 
which has holononmy group $SU(2) \times Sp(2 (N_V^{(A)} +1))$
and the appropriate negative curvature scalar. Likewise one can
explicitly find the three complex structures. We will work this
out for the case of one hyper multiplet in the next subsection and
refer to \hcite{ferr} for the general case. 

Moreover the action \href{0130} can be used to identify certain
interesting subspaces of the quaternionic manifold.
First one can set $C_I$ to constant, purely imaginary values,
$\p_{\mu} C_I=0$, $(C+\ov{C})_I=0$. Then the action reduces to
\be0140
S[z^A,S'] = \int d^4 x \sqrt{G^{IIB}} \left\{
-2 g_{A\ov{B}} \p z^A \p \ov{z^B} 
-2 \frac{\p_{\mu} S' \p^{\mu} \ov{S'}}{ (S' + \ov{S'})^2 }
\right\}
\eq
which has the K\"ahler target space $K_{N_V}^{(A)} \times
\frac{SU(1,1)}{U(1)}$ parametrized by $z^A$ and $S'$ 
respectively \hcite{ferr}. Note that $K_{N_V}^{(A)}$ is the dual
special K\"ahler manifold. On the other hand one can
set $z^A$ to constant values and explore the directions that
have been added to $K_{N_V}^{(A)}$ by the s-map. 
After shifting
\be0150
S' \longrightarrow 
S = S' - \frac{1}{2} C_I (\Im {\cal N})^{IJ} C_J\,,
\eq
ones gets the action 
\be0160
S[S,C_I] = \int d^4x \sqrt{G^{IIB}} \left\{ 
- 2 \wt{K}_{S\ov{S}} \p_{\mu} S 
\p^{\mu} \ov{S}
- 2\wt{K}_{S \ov{C_J}}  \p_{\mu} S \p^{\mu} \ov{C_J}
- 2\wt{K}_{C_I \ov{S}} \p_{\mu} C_I \p^{\mu} \ov{S}
- 2\wt{K}_{C_I \ov{C_{J}}} \p_{\mu} C_I 
\p^{\mu} \ov{C_J} \right\}\,,
\eq
with $\wt{K} = - \log( S+\ov{S} - C_I \Im {\cal N}^{IJ} \ov{C_J})$.
This shows that $S$ and $C_I$ parametrize for fixed $z^A$
the K\"ahler manifold \hcite{ferr}
\be0170
K(S,C_I) =
\frac{SU(1,N_V^{(A)} +2)}{U(1) \times 
SU(N_V^{(A)} +2)} \subset Q^{(B)}_{N_V^{(A)}+1}
\eq
Thus the directions added to 
$K_{N_V^{(A)}}$ by the s-map are completely universal.


By construction, special quaternionic manifolds have a lot of
isometries. The gauge symmetries of the field strength $F^I_{\mu \nu}$ and
of the Kaluza Klein boson $\omega_m$ imply that the scalars
$\zeta^I$, $\wt{\zeta}_I$ and $\wt{\phi}$ have axion-like shift
symmetries. Together with a scale symmetry of $\phi$ this gives
a total of $2N_V^{(A)} + 4$ isometries, that act on the complex
fields as \hcite{ferr}
\bea0180
 S' &\rightarrow& S' + i \alpha -2 C_{I}\gamma^{I} 
                 - \gamma^{I}{\cal{N}}_{IJ}\gamma^{J} 
\nonumber\\
 S' &\rightarrow& \lambda \ S'
\nonumber\\
 C_{I} &\rightarrow& C_{I} + i \beta_{I} + 
                           {\cal{N}}_{IJ}\gamma^{J} 
\nonumber\\
 C_{I} &\rightarrow& \lambda^{1/2} \ C_{I}
\eqa
with $2N_V^{(A)} +4$ real parameters $\alpha$, $\beta_I$, $\gamma^I$ and
$\lambda$. Due to these isometries, the quaternionic metric does only
depend on $z^A$ and $\Re S$ (or $z^A$ and $\phi$). Note that
this is only the minimal set of isometries. In addition, all isometries
of the dual special K\"ahler manifold are isometries of the quaternionic
manifold, and there may be further isometries as well \hcite{WVV}.

\subsection{The universal sector}

The c-map involves both space-time quantities (metric, graviphoton,
dilaton) and quantities related to the internal Calabi-Yau threefold
(K\"ahler and complex structure moduli) and mixes them. In order
to separate these two kinds of quantities and to have a simple
and illustrative example it is useful to consider 
the minimal case in which the c-map makes sense, namely 
$N_V^{(A)} = 0$, $N_H^{(A)} = 1$, which is called $N=2$
dilaton-supergravity. Before going into this let us note
that this case cannot be obtained by Calabi-Yau compactification
because a Calabi-Yau threefold has at least one modulus (its 
K\"ahler class, i.e. its overall radius) implying that one has
at least one additonal vector or hyper multiplet. Nevertheless
dilaton-supergravity appears as a subsector in every
Calabi-Yau compactification, characterized by setting all 
multiplets related to internal degrees of freedom to zero
(and hence only depending on space-time quantities).
Note that this way one can in particular embedd all solutions of
dilaton-supergravity into every Calabi-Yau compactification.

{From} the discussion in the last subsection it is clear that
dilaton-supergravity is selfdual under the c-map, which
simply exchanges the gravity multiplet (containing the graviton and
the graviphoton as its bosonic part) with the single hyper multiplet
(containing 4 scalars including the dilaton).

The scalar part of the action can be obtained by 
setting $z^A$ = const. and $C_A$ = const. in (\href{0084})
or more simply by setting 
$C_A =$ constant in (\href{0160}). Defining
$C= \Im ({\cal N}^{-1/2})^{0I} C_I = \Im ({\cal N}^{-1/2})^{00} C_0$
we get a K\"ahler $\sigma$-model with K\"ahler potential
$K=-\log( S + \ov{S} - C\ov{C})$, i.e. the moduli space associated
with the universal hyper multiplet is
\be0190
{\cal M}(S,C) = \frac{SU(2,1)}{SU(2) \times U(1)}
\eq
Thus in this case the hyper multiplet moduli space is K\"ahler.

In order to display the quaternionic structure it is useful to work with
the shifted field $S'$ instead of $S$, because then the 
action can be written in terms of holomorphic squares of 
$\p_{\mu} S'$ and $\p_{\mu} C$ \cite{ferr}. {From} \href{0130}
one obtains
\be0200
S^{H} = \int d^4 x \sqrt{G} \left\{ -2 e^{2 \wt{K}}
|\p_{\mu} S' - (C+\ov{C}) \p_{\mu} C |^2
-2 e^{\wt{K}} |\p_{\mu} C|^2 \right\}
\eq
with $\wt{K} = - \log (S' + \bar S' - \frac{1}{2} (C + \bar C)^2)$.
Introducing the complex 1-forms
\be0210
u =  e^{\tilde K/2} \ dC\,,  \;\;\;
v = e^{\tilde K} \ dS' - e^{\tilde K} \ (C + \bar C) dC 
\eq
we can rewrite (\href{0210}) as
\be0220
S[S',C] 
=\int \left\{ - u \wedge \star \ov{u} - v \wedge \star \ov{v} \right\} \,.
\eq
We introduce the quaternionic vielbein 
\be0225
V = \left( \begin{array}{cc}
u & \ov{v} \\  v & - \ov{u} \\
\end{array} \right) = i\, \Im (u) \, {\bf 1} + \Re (v) \, \sigma_1 
+ \Im (v) \, \sigma_2 + \Re (u) \, \sigma_3 \,.
\eq
(Recall that the quaternionic algebra is isomorphic to the one 
generated by ${\bf 1}$ and $-i \sigma_i$.)
The matrix indices labeling the rows and columns are $SU(2)$
indices and refer to the first (second)
factor of the holonomy group $SU(2) \times SU(2)$ as we will see in
a moment. Using 
\be0230
 du = -\frac{1}{2} (v + \bar v) \wedge u  \;\;\;\mbox{and} \;\;\;
 dv =  v \wedge \bar v + u \wedge \bar u 
\eq
the connection $\Omega$ can be found from the covariantly constancy
condition,
\bea0240
 ( d \ + \ \Omega ) \ V &=& 0 \,,
\end{eqnarray} 
with the result
\be0232
\Omega = p \otimes {\bf 1}_{2} +
{\bf 1}_{2} \otimes q\,,
\eq
where
\be0234
p = \left( \begin{array}{cc}
\frac{1}{4} ( v - \ov{v} ) & -u \\
\ov{u} & - \frac{1}{4}(v - \ov{v}) \end{array} \right)
\mbox{   and   }
q  = \left( \begin{array}{cc}
- \frac{3}{4} ( v - \ov{v} ) & 0 \\
0 & \frac{3}{4} ( v - \ov{v} ) \\
\end{array} \right) \,.
\eq
The resulting curvature is
\be0236
R = d \Omega + \Omega \wedge \Omega
= R_1 \otimes {\bf 1}_{2} + {\bf 1}_{2} \otimes R_2
\eq
where $R_i$ are $SU(2)$ curvatures. The curvature of the first factor
is related to the hyper K\"ahler form $J$:
\be0238
R_1 = dp + p \wedge p = \left( \begin{array}{cc}
\frac{1}{2} ( \ov{u} \wedge u - \ov{v} \wedge v)&
\ov{v} \wedge u \\
\ov{u} \wedge v & 
-\frac{1}{2} ( \ov{u} \wedge u - \ov{v} \wedge v)\\
\end{array} \right) = \sum_{i=1}^3 \frac{1}{2} \alpha_i \sigma^i 
=-iJ
\eq
where the expansion coefficients $\alpha_i = \ov{w^a} \sigma^i_{ab}
w^b$, $(w^a)= (u,v)^T$ are explicitly given by
\be0239
\alpha_1 = \ov{u} \wedge v + \ov{v} \wedge u, \;\;\;
\alpha_2 =
-i \ov{u} \wedge v + i \ov{v} \wedge u, \;\;\;
\alpha_3 = \ov{u} \wedge u -  \ov{v} \wedge v \,.
\eq
The three complex structures $J_i$ are found by converting the 2-form
components $\alpha_i$ of the hyper K\"ahler form $J$ into 
$(1,1)$ tensors 
$J_i = -i \ov{w^{\star a}} \sigma^i_{ab} w^b$,
where $\ov{w^{\star a}}$ is the vector field dual to the 1-form $w^a$.
The required  quaternionic algebra (\href{0008})
is obtained with 
$\ov{u^{\star}} \wedge u + \ov{v^{\star}} \wedge v$ as
its unit element. 
Thus we have made the quaternionic structure obvious by
identifying the hyper K\"ahler form $J$ and the three complex
structures $J_i$, which are by construction covariantly constant with
respect to the $SU(2)$ connection. 

Finally we have to check the constraint on the curvature scalar.
The curvature of the second $SU(2)$ factor is
\be0242
R_2 = dq + q \wedge q = - \frac{3}{2} 
(v \wedge \ov{v} + u \wedge \ov{u}) 
\left( \begin{array}{cc} 1&0\\0&-1\\ \end{array} \right)
\eq
Computing the Ricci-scalar of the full curvature one finally finds
${\cal R}(\Omega)=-24$, which according to (\href{0009}) is the
correct result for $N_H=1$.
This completes the proof that the scalar manifold of
$N=2$ dilaton-supergravity (\href{0190}) is a quaternionic manifold that
is consistent with local $N=2$ supersymmtery \hcite{ba/wi}.

There are two different ways to break the moduli space
of the universal sector (\ref{0190}) down to 
the coset space $\frac{SU(1,1)}{U(1)}$.  
First, one neglects all R-R scalars, setting
$S'=\phi + i \wt{\phi}$, $C=0$:
\bea0310
S[C] &=& \int d^4 x \sqrt{G} \left\{
-2 \frac{ |\p_{\mu} S'|^2 }{ (S' + \ov{S'})^2 } \right\} =
\int d^4 x \sqrt{G} \left\{
-2 \frac{ (\p_{\mu} \phi)^2 + (\p_{\mu} \wt{\phi})^2 }{ 4 \phi^2 }
\right\} \nonumber \\
 &=&
\int d^4 x \sqrt{G} \left\{
-2 \left(  (\p_{\mu} \varphi)^2 + \frac{1}{4} e^{4 \varphi} 
(\p_{\mu} \wt{\phi})^2 \right) \right\} \\
\nonumber
\eqa
where we introduced the standard four-dimensional dilaton
$\varphi$ by $\phi = e^{-2 \varphi}$
for comparison with section 2.
This yields the standard NS-NS dilaton-axion moduli space. 

Second, one keeps the dilaton together with
one of the R-R scalars, which for definiteness we take to be the
R-R pseudoscalar $\wt{\zeta}_0$. Setting $S' = \phi$ and 
$C = i \wt{\zeta}_0$ gives 
\bea0320
S[T] &=& \int d^4 x \sqrt{G} \left\{
-2 \frac{ |\p_{\mu} T|^2 }{ (T  + \ov{T})^2 } \right\} =
\int d^4 x \sqrt{G} \left\{
-2 \left( \frac{ (\p_{\mu} \phi)^2 }{ 4 \phi^2 }
+ \frac{ (\p_{\mu} \wt{\zeta}_0)^2 }{ 2 \phi }
\right) \right\} \nonumber \\
& =&
\int d^4 x \sqrt{G} \left\{
-2 \left(  (\p_{\mu} \varphi)^2 + \frac{1}{2} e^{2 \varphi} 
(\p_{\mu} \wt{\zeta}_0)^2 \right) \right\} \\
\nonumber
\eqa
where $T = \phi + i a$ with $ (\p_{\mu} a)^2 / 2 \phi = 
(\p_{\mu} \wt{\zeta}_0 )^2$.
Comparing (\href{0320}) to (\href{0310}) we explicitly see that the
R-R scalar $\wt{\zeta}_0$ couples to the dilaton in a different way
than the NS-NS scalar $\wt{\phi}$. The action (\href{0320}) is the
natural starting point for the construction of $D=4$, $N=2$
D-instanton solutions.


\section{Stationary IIA solutions 
and IIB D-instantons}
\setcounter{equation}{0}

We now turn to the construction of explicit solutions
of the equations of motion (\href{0020}) - (\href{0060}).
During the last year a lot of non-trivial static and 
more recently stationary solutions to the 
gravity and vector multiplet equations have been found
while keeping the hyper multiplet scalars constant, so that 
(\href{0030}) is trivially solved. These solutions
describe in the static case charged black holes 
with various moduli fields turned on. Using T-duality and
the c-map we can relate any such solution to a 
configuration with trivial (Euclidean, Einstein frame)
metric and trivial vector multiplets, but non-trivial
hyper multiplet scalars. These solutions are instantons 
and contain as a subclass four-dimensional D-instanton
solutions, which in addition to a non-trivial dilaton
have several non-trivial moduli fields.

For the stability of solutions it is crucial to have 
BPS saturated solutions which are invariant under at least
one supersymmetry transformation. 
This is the case for all
the configurations that we discussed in section 2. 
Our general strategy here will be to start with supersymmetric
solutions on the IIA side and to apply T-duality in order to
get a solution on the IIB side. Moreover these solutions can be 
interpreted as Calabi-Yau compactifications of supersymmetric
D-$p$-brane solutions.
Therefore we expect that our solutions
are supersymmetric. A detailed check is postponed to future work
together with a more detailed analysis of our solutions.

The plan of this section, which contains our main 
results, is as follows. In section 4.1 we
study dilaton supergravity, i.e. the gravity multiplet
together with the universal hyper multiplet. In 
4.1.1 we first recall the extreme Reissner-Nordstrom black
hole, which is a BPS solution of pure $N=2$ supergravity.
Then we find the T-dual D-instanton solution by solving
the equations of motion of the T-dualized action using
an instanton ansatz analogue to
\hcite{GibGrePer}. We explore the
geometry of the D-instanton, which is a finite wormhole
in the string frame and compare it to its ten-dimensional
analogue and to the Reissner-Nordstrom black hole.  
Some related configurations, like NS-NS instantons are
briefly mentioned. In 4.1.2 we discuss the D-instanton
in the context of the quaternionic geometry of the 
hyper multiplet moduli space.

Section 4.2. is devoted to the IIB duals of 
non-axionic IIA black holes. If the IIA black hole is in addition
double-extreme, then the non-trivial IIB fields live in a 
symmetric K\"ahlerian subspace of the quaternionic hyper multiplet moduli
space. In this case one can work on the IIB side with complex fields
which are standard coset coordinates. This is used in 4.2.1 to explore
this part of the moduli space. In particular we find that non-axionic
solutions can be characterized by a reality constraint. When moving
away from the double extreme limit on the IIA side in 4.2.2,
it is more convenient
to work in terms of real fields on the IIB side. Then
the solution still looks simple and its structure is very similar
to the well known non-axionic IIA black holes. In particular
the solutions can be expressed in terms of harmonic functions.
At the end of 4.2.2 we interpret the solutions from the ten-dimensional
point of view, compare them to those obtained in section 2 by toroidal
compactification and comment on the role of the geometry of the internal
Calabi-Yau threefold.

Section 4.3 gives our central result, an explicit set of formulae which
specifies the fact that instantons are T-dual to the most general 
stationary solution of the gravity/vector multiplet
sector. The results of section 4.1 and 4.2 should
be considered as first and relatively simple application
of this formalism.

\subsection{Dilaton supergravity}
\subsubsection{RN black holes in pure supergravity and  D-instantons in D=4}
\setcounter{equation}{0}

In this section, we discuss explicit solutions in $D = 4$ and the 
associated c-map. First, we relate the black hole solution on the type 
IIA side and the D-instantons of type IIB theory through the
c-map in $D=4$. We take $G_N = 1$.  

The bosonic part of the action of ungauged $N=2$ supergravity coupled to 
$N_V$ vector multiplets is given by,
\bea2001
e^{-1} {\cal L} &=& R - 2 g_{A \bar B} 
                    \partial_{\mu} z^A  \partial^{\mu} \bar z^{\bar B} - 
\frac{1}{4} {\mbox{Im}} \ {\cal N}_{IJ} \ F_{\mu\nu}^{I} F^{J\mu\nu} - 
\frac{1}{4} {\mbox{Re}} \ {\cal N}_{IJ} \ F_{\mu\nu}^I {}^{\star} F^{J\mu\nu}
\end{eqnarray}

Here the index $I$ runs from $0, 1, \ldots ,
 N_V$, where the extra index 
0 is due to the graviphoton degrees of freedom. 
The D-instanton of \cite{GibGrePer} contains the 
dilaton and the
R-R scalar as non-trivial fields. They are world-sheet 
configurations which correspond to a target space-time event giving 
rise to exponentially suppressed contributions to scattering amplitudes
of order $e^{-\frac{1}{g}}$, where $g$ denotes the string coupling constant.
The dilaton has its origin in
the dual type IIA metric and the RR scalar in the graviphoton.
The c-map maps these two fields to the dilaton and the RR scalar
on the type IIB side respectively. Thus, it is sufficient to
consider pure $N=2$ (type IIA) supergravity containing only the 
graviphoton and we go to the universal
IIB hyper multiplet via c-map. 
The prepotential for pure $N=2$ supergravity is given by, 
$F(X)=2 i (X^0)^2$. We show here that the RN black holes in pure 
supergravity are related to the D-instantons via c-map. It is sufficient
in the following to discuss only the electric type black hole 
solution on the type 
IIA side.  
It follows from the previous definition that,
${\cal N}_{IJ}={\cal N}_{00}=i$. Thus, the Lagrangian  
is given by 
\bea2002
e^{-1} {\cal L} &=& R -
\frac{1}{4} \ F_{\mu\nu} F^{\mu\nu}.
\end{eqnarray}
The equations of motion are obtained as, 
\bea2003
\nabla_{\mu} \ F^{\mu\nu} &=&  0 \\
R_{\mu\nu} + \frac{1}{8} \ g_{\mu\nu} F_{\sigma\rho} F^{\sigma\rho} 
+ \frac{1}{2} F_{\mu\rho} F^{\rho}_{ \ \ \nu}&=& 0.
\end{eqnarray}
The corresponding spherically symmetric black hole solution is given 
by the well-known Reissner-Nordstrom solution.
In the Einstein frame, the metric
has the following particular form in Minkowski space ($x^{\mu}=x^0,x^m$)
\bea2004
ds^{2}_{E} &=& - \ e^{2 U(r)} \ dt^2 +  \ e^{-2 U(r)} \ d {\vec x}^2
\end{eqnarray}
Note that the real dilaton in the type IIA
model is trivial, but the metric is not flat. What we call here
$U$ is related to the dilaton on the type IIB side ($U=-\varphi$)
The components of the Ricci tensor are given by,
\bea2005
R_{mn} &=&  g_{mn} \partial^2 U - 2  \partial_m U \partial_n U
\\
R_{00} &=& - g_{00} \ \partial^2 U 
\end{eqnarray}
Then the Ricci scalar reads as,
\bea2006
R &=& 2 \partial^2 U - 2 (\partial U)^2 
\end{eqnarray}

Restricting ourselves to the pure electric solution 
$F_{\mu\nu} \sim F_{0m}$ we find three equations of motion, namely,
\bea2007
\partial_m (e^{-2U} F^{m0}) &=& 0
\\
R_{00} - \frac{1}{8} g_{00} F^2 &=& 0
\\
R_{mn} + \frac{1}{8} g_{mn} F^2 - \frac{1}{2} F_{m0} \ g^{00} \ F_{n0} &=& 0
\end{eqnarray}  
Taking $F_{m0} = \partial_m A$
it follows that $F^2 = -2 (\hat\partial A)^2$, where
hatted quantities refer to flat space 
($\hat\partial^2=\eta^{mn}\partial_m\partial_n$).
Using the ansatz $A=2 e^U$, we find that the equations of motion
are satisfied, provided    
\bea2008
\hat\partial^2 \ e^{-U} &=& 0.
\end{eqnarray}
Thus we find that the pure electric Reissner Nordstrom black hole
is specified by the harmonic function
\bea2009
e^{-U(r)} &=&  e^{-U_{\infty}} \ + \ \frac{q_0}{r}.
\end{eqnarray}
In the following we assume flat space at infinity
($ e^{-U_{\infty}}=1$).

{From} the spatial component of the metric we can read off the
entropy and the ADM mass, given by the expressions,
\be2011
\label{bh1}
S_{BH} = \pi (r^2 g_{rr})_{r=0}     
= \pi \ q_0^2,  \hspace{2cm}
M_{ADM} = q_0  
\eeq
On the other hand, recent developments in four dimensional
$N=2$ supergravity and string theory
show that the entropy and the mass of a 
BPS saturated black hole are completely determined by the
corresponding prepotential
\bea2012
\label{bh2}
S_{BH} &=& \pi \  |Z|^2,  
\hspace{2cm}
M_{BPS}^2 \ = \  |Z|^2 \ = \ e^{K(z,\bar z)} |q_I X^I(z)-p^I F_I(z)|^2.
\end{eqnarray}
where $Z$ is the central charge, $q_I$ and $p_I$ are the electric 
and magnetic charges respectively. It is easy to verify that, 
for the case at hand, 
(\ref{bh1}) and (\ref{bh2}) agree with each
other. Hence, the corresponding electric Reissner Nordstrom black
hole preserves one half of the $N=2$ supersymmetry.

Now we discuss the corresponding picture on the type IIB side. The Lagrangians
in both theories are dual to each other as discussed before and 
are related by the c-map. Performing the c-map, the dual Lagrangian 
in terms of real fields is given by, 
\footnote{in comparison with \cite{ferr},
we have  $\tilde\zeta= 0 = \tilde\phi$,
$\phi = e^{-2 \varphi}$.}
\bea2013
\label{euclid_einstein_action}
e^{-1}{\cal L} &=& 
R -  2 (\partial \varphi)^{2}  
-  e^{2\varphi} (\partial\zeta^0)^{2}.
\end{eqnarray}
The scalar fields parametrize the $\frac{SU(1,1)}{U(1)}$ coset.
Since the c-map exchanges the $N=2$ supergravity sector with
the type II dilaton in the universal sector, 
the metric is now flat while the
type IIB dilaton is non-trivial. The Euclidean version of this action 
represents the action corresponding to D-instanton in type IIB 
theory. 
In 4-dimensions,
one has the duality between the D-instanton ($p = - 1$
brane) and D-string ($p = 1$ brane) analogous to the D-instanton and 
$7$-brane duality in ten dimensions. 
One could also have started with a magnetically charged solution in the
type IIA side, for which $F_{0 m} = 0$ and $F_{m n} \sim 
\epsilon_{m n p} \partial\tilde
\psi^0$, where the field $\tilde\psi$ is later identified with 
$\tilde\zeta$
in comparison with \cite{ferr}. 
This is because of the electric-magnetic duality in 4 
dimensions where the black hole in general is dyonic as the dual of a
2-form gauge field strength is also a 2-form in $D=4$. 

In order to discuss the D-instanton in type IIB theory, we perform a 
Wick rotation of $\z^0 \rightarrow i \z^0$ to go to the Euclidean 
version. 
The Euclidean action is given by,
\be2099
e^{-1}{\cal L}_E = R - 2(\p\varphi)^2 + e^{2\varphi} (\p\zeta^0)^2
+ \mbox{boundary terms} 
\eq
The origin of the boundary term has been discussed in section 2 and we 
shall comment on it later. The corresponding field equations in 
Euclidean spacetime read as,
\bea2014
\label{eq_motion}
\nabla_{\mu} \ ( \ e^{2\varphi} \ \partial^{\mu} \zeta^0 \ ) &=&  0
\\
2 \nabla^2 \varphi \ + \ e^{2\varphi} \ (\partial\zeta^0)^{2} &=&  0 
\\
R_{\mu\nu} 
\ - \ 2 \partial_{\mu} \varphi \ \partial_{\nu}  \varphi
\ + \ e^{2\varphi}\partial_{\mu}\zeta^0  \partial_{\nu} \zeta^0
&=& 0
\end{eqnarray}

Now we consider the instanton ansatz of \hcite{GibGrePer} with
$d\zeta^0 = {\sqrt 2} e^{-\varphi} d\varphi$ and flat Euclidean metric
$g_{\mu\nu} = \delta_{\mu\nu}$ in our four dimensional context.
Note that the fields still only depend on 3 spatial coordinates and
are independent of the time $t$.
With this ansatz, one gets the constraint $\partial^2\varphi = - (\partial
\varphi)^2$ and the instanton ansatz leads to the equation of motion, 
\bea2015
\hat\partial^{2} e^{\varphi} =  0
\end{eqnarray}
with general 3-dimensional spherically symmetric solution
\bea2016
\label{eq_motion_solution}
 e^{\varphi} &=&  e^{\varphi_{\infty}} \ + \ \frac{q_0}{r} 
\end{eqnarray}

The string and Einstein frame metrics associated with this solution 
are:
\be2088
ds^2_S = e^{2 \varphi}d s^2_E = (e^{\varphi_{\infty}} + \frac{q_0}{r})^2
(d t^2 + d r^2 + r^2 d\Omega_2^2)
\eq
with $r^2 = x^2 + y^2 + z^2$. The $t$ direction is an isometry, 
whereas the geometry in the three other directions is an infinite throat 
interpolating between the flat two-space and a two-sphere of radius 
$|q_0|$. This configuration is T-dual to the Euclideanized extreme 
Reissner-Nordstrom black hole. Now we can relax the constraint 
that the solution 
has the isometry in the Euclidean time direction and allow for a more 
general dependence on all four coordinates. So the general spherically 
symmetric solution is given by,
\be2087
e^{\varphi} = e^{\varphi_{\infty}} + \frac{q_0}{r^2}
\eq
This is the D-instanton solution in four dimensions. The dilaton is 
singular at the origin and the Einstein metric is flat.  
Here $\varphi_{\infty}$ is the value of the dilaton at infinity, which we 
take to be constant. On the other hand, if we transform the instanton
solution to the string frame with the metric $G_{\mu\nu} = 
e^{2\varphi} g_{\mu\nu}$, then  
we find that the D-instanton solution 
\bea2017
 ds^{2}_{S} &=&  ( e^{\varphi_{\infty}} \ + \ \frac{q_0}{r^2} )^{2} \
 (d r^2 + r^2 d\Omega_3^2)
\end{eqnarray}
is invariant under the transformation,
\be2086
r \,\rightarrow \frac{q_0 e^{- \varphi_{\infty}}}{r}
\eq

The interpretation of this result is as follows: in the string frame 
the instanton solution is a wormhole solution connecting two asymptotically
flat Euclidean regions by a neck. The D-instanton solution on the 
IIB side is expected to be supersymmetric as the corresponding black
hole solution in the IIA theory is supersymmetric.   

The electric Noether charge $q_0$ associated with the Noether current
$j_\mu = e^{2\varphi} \partial_\mu A$
of the global (Peccei-Quinn) symmetry transformation 
$\zeta \rightarrow \zeta + \theta$ with $\theta \in {\bf R}$ reads as,
\bea2018
\label{electric_charge}
\oint_{\partial M}  
j_\mu \ d\Sigma^{\mu} &=& 
- \oint_{\partial M} \hat\partial_\mu e^\varphi d\Sigma^{\mu}
\ = \ 2 \pi q_0 
\end{eqnarray}
The corresponding D-instanton action comes from the
boundary only. The action in the bulk vanishes by substituting the 
instanton ansatz. Contribution from the boundary is given by,
$S_{\partial M} = - 4 \partial^2\varphi$. Hence the instanton action is
given by,
\bea2019
 S_{\rm inst} &=& 
- 4 \int_{M^4} d^4x \ \hat\partial^2 \varphi \ = \
- 4 \oint_{\partial M^2}d\Sigma_{\mu} \ \hat\partial^\mu \varphi 
\ = \ 8 \pi \frac{|q_0|}{g}. 
\end{eqnarray}
{From} the point of view of the dual type IIA model the type IIB
D-instanton action has its origin in the four-dimensional Einstein
action and T-duality. This is an interesting observation with 
respect to the uncompactified ten-dimensional type IIB theory.
There the D-instanton action has been derived using D-instanton /
D-7-brane Poincare duality \cite{GibGrePer,GreGut}. In our
four-dimensional context this corresponds to the Poincare duality
of a D-string and a D-instanton. In the context of the
D-string / D-instanton duality
one can start with the Minkowski action written in terms 
of the hodge dual three form field strength and then perform a Wick 
rotation to 
obtain the Euclidean action which is manifestly positive as the $F^2$ 
term does not change sign under Wick rotation and one does not need to 
introduce a boundary term. On the other hand, if we dualize the three form
action to scalar(pseudoscalar) field action, then one has to take proper 
consideration of a boundary term, which makes the scalar action positive. 
To obtain the scalar action with proper boundary term, one starts with 
the action in terms of a 3-form field strength and adds a Lagrange multiplier
term involving the three form and the 0-form. Integrating over the 
0-form (scalar field), one obtains the dual action for the 3-form. To
go to the Euclidean version, one 
uses the properties of $p$-forms under
Poincare duality in 4 dimensions. Integrating over the 3-form, one obtains
the action for the scalar field together with the required boundary term.  
The presence of the boundary term is really crucial in computing the action 
in the instanton background. This is also true for D-instantons in ten 
dimensions \cite{GibGrePer,GreGut}. 

Since we started our analysis on the type IIA side, which was
determined by the $N=2$ supergravity prepotential only, the dual
type IIB D-instanton action is determined by the 
dual type IIA model. Moreover, in terms of the central charge
of the type IIA model we find 
$S_{\rm inst, IIB}\sim |Z|_{IIA}$.

It is interesting to consider what happens if we replace the R-R 
scalar $\zeta^0$ by the NS-NS scalar $\tilde\phi$ which has a 
different coupling to the dilaton. In this case, the instanton 
ansatz goes through with the simple modification that now $e^{2\varphi}$
has to be harmonic. Taking this function to depend on the 4-radius
$r$, where $r^2 = t^2 + x^2 + y^2 + z^2$, the string and Einstein 
frame metrics are related by, 
\beq
ds_S^2 = e^{2\varphi} ds_E^2 = (e^{2\varphi_{\infty}} + \frac{q_0}{r^2})
(d r^2 + r^2 d\Omega_3^2)
\eeq
This describes a semi-infinite wormhole which interpolates between 
the flat 3-space at $r\rightarrow\infty$ and a 3-sphere of radius 
$|q_0|$ at $r\rightarrow 0$ \cite{Callan}.
This time the instanton action is 
proportional to $\frac{1}{g^2}$. By restricting the dependence of 
$\varphi$ such that one gets an isometry, one can further T-dualize
this solution into a 0-brane type solution. Clearly this result 
will be different from the extreme Reissner-Nordstrom solution, since 
in the expression for the metric, the harmonic function appears with a 
first power instead of a second power. But this is to be expected because
the scalar $\tilde\phi$ comes from dualizing the Kaluza-Klein vector 
$w_m$. So it is a priori clear that the T-dual 0-brane is a non-static
stationary solution.

\subsubsection{D-Instantons and quaternionic structure}

In the last section, we discussed about the dualized black hole
solution in type IIA side and showed how to obtain the D-instanton 
solution in the type IIB theory by using the c-map. Here 
we explore the quaternionic geometry associated with the  
$D$-instantons.
We consider the minimal case of the type IIB dilaton
hyper multiplet coupled to $N=2$ supergravity. 
Here, we consider the case where there is no vector multiplet. 
So the four real scalars 
namely $\phi$, $\tilde\phi$, $\zeta^0$ and $\tilde\zeta_0$ 
of the quaternion can be
combined into 2 complex scalars $S'$ and $C^I$ as before, but in this 
case, the index runs only over $0$ as we are in the pure gravity 
sector. 
For the D-instanton, the $S'$ and $C$ fields in terms of non vanishing
scalars are given by the expression,
\bea2020
S' &=& \phi + (\zeta_0)^2\\
C_0 &=& - \zeta_0 \\
\nonumber
\eeqa
which follows from the general definition of the complex fields 
$S'$, $C_I$ and by putting $\tilde\phi = 0 = \tilde\zeta$. The action 
in terms of the complex fields looks like,
\bea2021
S^{H}_{II B} &=& \int d^4 x \sqrt{G^{II B}} \left\{
- 2 \frac{{|\partial S' - (C_0 + \bar C_0)\partial C_0|}^2}
{(S' + \bar S' - \frac{1}{2} (C_0 + \bar C_0)^2)^2} - 2 \frac{{|
\partial C_0|}^2}{(S' + \bar S' - \frac{1}{2} (C_0 + \bar C_0)^2)}
\right\} \\
\nonumber \eeqa
This is equivalent to the dualized type II A action written before.
In order to display the coset structure parametrized 
by the complex fields, we shift the $S'$ field as
\be2022
S' \rightarrow S = S' - \frac{1}{2} C_0^2
\eq
and the Lagrangian can be written as, 
\be2023
e^{-1} {\cal L} = - 2 g_{a \bar b} \partial_{\mu} z^a 
\partial^{\mu} {\bar z}^b 
\eq
where, $g_{a\bar b} = \p_a \p_{\bar b}\wt K$ and $\widetilde K$ is 
the K\"ahler metric given by the expression, 
\be2024
\widetilde K = - \log (S + \bar S - C_0 \bar C_0)
\eq
The complex fields $S$ and $C_0$ parametrize the K\"ahler manifold 
${\cal M} = \frac{SU(1, 2)}{U(1)\times SU(2)}$ and the coset metric 
is given by,

\be2025
{\bf g} = (S + \bar S - C_0 \bar C_0)^{-2}
\left( \begin{array}{cc}
1 & - C_0 \\ - {\bar C_0} & S + \bar S \\
\end{array} \right)
\eq

Next, we show that the moduli space of the $N = 2$ supergravity
is a quaternionic manifold, ${\rm i. e.}$ we consider the minimal case of 
the type II B dilaton hyper multiplet coupled to $N = 2$ supergravity.
Here, the 4 real scalars $\phi$, $\tilde\phi$, $\zeta$ and $\tilde\zeta$
(which comprise the universal hyper multiplet) are again combined
into the two complex fields,
\bea2026
S' &=& \phi + (\zeta_0)^2 + i \tilde\phi - \zeta^0 \tilde\zeta_0 \\
C_0 &=& - \zeta_0 + i \tilde\zeta_0 \\
\nonumber
\eeqa
Now we introduce the two one-forms namely,
\be2027
u = e^{\frac{\tilde K}{2}} d C; \, \, \,  v = e^{\tilde K} (dS' - 
(C + \bar C)dC)
\eq 
with $\tilde K$ as defined before. With 
this definition, the IIB Lagrangian becomes,
\be2028
- e^{- 1} {\cal L}_0 = (u, \bar u) + (v, \bar v)
\eq
where, $(,)$ denotes contraction of components. Here, we have $d{\cal N}
= 0$, that is the type II A dilaton is constant. The torsion of the 
quaternionic vielbein is given by,
\be2029
\left( \begin{array}{cc} du \\ dv\\
\end{array} \right) = \left( \begin{array}{cc}
- \frac{1}{2}(v + \bar v)\wedge u \\
v\wedge\bar v + u\wedge\bar u \\
\end{array}\right)
\eq

The curvature 2-form $R$ can be directly computed and one can obtain 
the expression for Hyper-\Kahler  2-form associated with the $SU(2)$
connection as discussed in the quaternionic geometry of the universal
hyper multiplet in section 3. So one obtains the result that the scalar
manifold of $N=2$ dilaton-supergravity is quaternionic in agreement 
with \cite{ferr}.


\subsection{Static solutions of ungauged $N=2$, $D=4$ Supergravity}

Here, we discuss IIB instanton solutions, which are T-dual
to specific static IIA solutions. 
The explicit solutions 
can be obtained as special cases of the general
dualized stationary IIA solution, which is presented 
in section 4.3. 

\subsubsection{Type IIB dual of non-axionic, double extreme II A 
black holes} 

In the last section, we dealt with the RN black hole solution and
its T-dual version, which was shown to correspond to the D-instanton
solution in type IIB theory. We discussed the simplest case of 
c-map involving $N_V^{(A)} = 0$ and $N_H^{(A)} = 1$.  
Here
we consider the more general case of $N_V$ vector multiplets 
coupled to $N = 2$ supergravity. So in the case, we have $N_V$ physical 
complex scalars $z^A$, the R-R scalars $C_I (I = 0,\ldots N_V)$ and
the complex field $S'$ which are the coordinates for the dual 
quaternionic manifold. While working out explicitly the geometry of 
the quaternionic metric of the IIB hyper multiplets, one finds that 
the computation becomes quite complicated as the derivative 
of the matrix ${\cal N} (z, \bar z)$ appears there and leads to 
various cross terms in the metric. If the matrix ${\cal N}$ becomes
holomorphic, then the quaternionic manifold becomes K\"ahlerian.
This can happen for the case of minimal coupling, {\it i.e.} quadratic 
prepotentials. But we do not consider this here as this case does 
not occur in string compactifications. However, we can consider
the case of type II A solutions with constant scalar fields 
{\it i.e.} $z^A = const$, which corresponds to double extreme black holes,
which are given by configurations where the moduli take constant 
values all the way from the horizon up to spatial infinity.
In this case, the Q manifold again becomes K\"ahlerian and is the 
manifold of pure R-R scalars. So the nontrivial scalars 
$S'$ and $C_I$ ($I = 0,\ldots N_V$) parametrize the K\"ahler quaternionic
manifold $\frac{SU(2 + N_V, 1)}{SU(2 + N_V)\otimes U(1)}$. In this case,
all the nontrivial scalars come from the IIA gravity multiplet and 
from the IIA field strength. 
The metric and the connections are sufficiently simple in order to 
explicitly write down the IIB scalar equation of motion, which are 
solved by the T-dualized double extreme black holes. On the IIB side, 
this corresponds to the generalizations of the D-instanton solutions
discussed in the previous section. 

In the following, we recall the standard parametrization of the 
coset, derive the metric and associated connection.  

The action in terms of $2(N_V^{(A)} + 1)$ complex scalars $S'$, $C_I$ 
and $z^A$ has been written down in section (3). 
For constant moduli fields (scalars $z^A$), the action reduces to,
\bea2040
S^H &=& \int d^4 x{\sqrt G}\{- 2 e^{2\wt K}|\p S' 
- (C + \ov C)_I (\Im{\cal N})
^{I J} \p C_J|^2 - 2 e^{\wt K}\p C_I (\Im{\cal N})^{I J}
\p \ov {C_J}\} \\
\nonumber
\eeqa
where the fields $S'$ and $C_I$ have been defined before in section 3. 
The coset 
structure becomes more transparent by rewriting the action in terms 
of the shifted field $S$. 
The K\"ahler potential $\wt K$ is given by, 
\be2041
\wt K = - \log(S + \bar S - C_I \Im {\cal N}^{I J}\ov {C_J})
\eq
In terms of canonical coset coordinates $S$ and $C$, the coset metric 
is given by,
\be2042
{\bf g} = (S + \bar S - C_I \Im {\cal N}^{I J}{\ov C_J})^{- 2}
\left( \begin{array}{cc}
1 & - C_J (\Im {\cal N}^{-\frac{1}{2}})_{J K} \\
- (\Im {\cal N}^{-\frac{1}{2}})_{I J} {\ov {C_J}} & 
\delta_{I K} (S + \bar S - C_J \ov {C_J}) + \ov {C_I} C_K \\ 
\end{array} \right)
\eq
The inverse metric is given by, 
\be2043
{\bf g}^{- 1} = (S + \bar S - C_I \Im {\cal N}^{IJ}\ov {C_I})
\left( \begin{array}{cc}
S + \bar S & C_J(\Im {\cal N}^{-\frac{1}{2}})_{JK} \\
(\Im {\cal N}^{-\frac{1}{2}})_{IJ} {\ov {C_J}} & \delta_{I K} \\
\end{array} \right)
\eq

Defining ${\cal C}_J = C_I(\Im {\cal N}^{-\frac{1}{2}})_{I J}$.
the connection as a matrix valued one-form is given by, ${\bf\Gamma}
= {\bf\Gamma}_S d S + {\bf\Gamma}_I  d {\cal C}_I$. The matrices 
${\bf\Gamma}_S$ and ${\bf\Gamma}_L$ are given in terms of matrices,
${\bf\Gamma}_S = {\bf g}^{-1}\p_S{\bf g}$ and ${\bf\Gamma}_L = 
{\bf g}^{-1}\p_L{\bf g}$. The explicit matrices are given by,
\be2044
{\bf\Gamma}_S = \frac{1}{S + \bar S - \C_J \bar \C_J}
\left(\begin{array}{cc}
- 2 & \C_K \\
0 & - \delta_{I K} \\
\end{array} \right)
\eq
and,
\be2045
{\bf\G}_L = \frac{1}{S + \bar S - \C_J \ov \C_J} \left(\begin{array}{cc}
2 \ov \C_L & -\ov \C_L \C_K - \delta_{K L}(S + \bar S - \C_J \ov \C_J)\\
0 & {\ov \C_L} \delta_{I K} \\
\end{array}\right)
\eq

The equation of motion of the real scalars $q_u$ are written down in 
section 3. If we consider $D$-instanton type solutions, then the space-
time metric is Euclidean and flat. If we further assume that the fields 
$q_u$ do depend on the 4-radius $r$, where $r^2 = 
t^2 + x^2 + y^2 + z^2$, 
then we get a geodesic equation in $r$,
\be2046
\frac{d^2 q_u}{d r^2} + \frac{3}{r}\frac{d q^u}{d r} + \G^u_{v w}
\frac{d q^v}{d r} \frac{d q^w}{d r} = 0 
\eq
We have previously combined the real scalars into complex ones, 
$S$ and $C_I$. 
The Christoffel connection coincides with the hermitean connection 
because the scalar manifold is \Kahler. 


Now, we consider dualizing an axion-free double extreme black hole 
solution with 
one electric and $N_V$ magnetic charges. For these black hole solutions,
we have,
\be2050
\Re {\cal N} = 0; \, \, \, \, {\cal N}_{0 A} = 0
\eq

Since the solution is static and has no magnetic components of 
$F_{\mu\nu}^0$ and no electric components of $F_{\mu\nu}^A$, it implies
that the following dual scalars vanish:
\be2051
\tilde\phi = 0;\, \, \, \tilde\zeta^0 = 0, \, \, \, \zeta_A = 0
\eq
{From} the above condition, it is clear that we are associating electric 
type solutions with $\z_0$ and magnetic type solutions with $\tilde\z^A$,
where $A = 1 \ldots N_V$. 

After substituting this in the hyper multiplet part of the dual action,
we obtain,
\be2052
S^H_{IIB} = \int d^4 x\,{\sqrt {G^{(IIB)}}}\left\{-\frac{1}{2\phi^2}(\p\phi)^2
- \frac{1}{\phi}\p\zeta^0 \Im {\cal N}_{00} \p\zeta^0 - \frac{1}
{\phi}\p\tilde\zeta_A \Im {\cal N}
^{AB}\p\tilde\zeta_B\right\}
\eq
Defining $e^{-2\varphi} = \phi$ as before, we rewrite the action as,
\be2062
S^H_{IIB} = \int d^4 x\,{\sqrt {G^{(IIB)}}}\left\{ -2 (\p\varphi)^2 - 
e^{2\varphi} \p \z^0 \Im {\cal N}_{00}\p\z^0 - e^{2\varphi} \p
\tilde \z_A \Im {\cal N}^{AB} \p\tilde \z_B\right\}
\eq
The Euclidean version of this action is a generalization of 
D-instanton action obtained 
previously by dualizing the black hole solution in pure gravity. The 
limit $\tilde\zeta_A = 0$ and $\Im {\cal N}_{00} = 1$, 
corresponds to D-instanton action in type IIB theory.
For the non-axionic case, the complex coordinates 
$S'$ and $C$ reduce to,
\bea2053
S'&=& \phi + \zeta^0\Im {\cal N}_{00}\zeta^0 \\
C_0 &=& - \Im {\cal N}_{00}\zeta^0; \\
C_A &=& i\tilde\zeta^A\\
\nonumber
\eeqa
Note that, $S'$ and $C_0$ are real while $C_A$ is imaginary. The 
\Kahler  potential is computed to be, $\wt K = - \log (2\phi)$.
The canonical coset coordinate $S$ is given by, 
\be2054
S = \phi + \frac{1}{2}\zeta^0 \Im {\cal N}_{00}\zeta^0 + \frac{1}{2}
\tilde\zeta_A \Im {\cal N}^{AB}\tilde\zeta_B
\eq

Thus using the c-map, the non-axionic double extreme back holes 
on the IIA side are 
mapped directly to the generalized D-instanton solutions in the IIB 
side which are defined by the reality constraints. The actions in both 
the cases are dual to each other. 

We also consider the case of double extreme black holes, where axions
are also allowed. In this case, the solution on the type IIA side 
is that of a stationary 
solution and using the c-map, the action in the dual theory in terms 
of non zero fields is given by,
\be2055
S^H_{IIB} = \int d^4 x{\sqrt {G^{(IIB)}}}\left\{-\frac{1}
{2\phi^2}(\p\phi)^2
- \frac{1}{2\phi^2} (\p\tilde\phi + \zeta^I\p\tilde\zeta_I - 
\p\zeta^I\tilde\zeta_I)^2 - \frac{1}{\phi}\p\zeta^I \Im {\cal N}_{IJ}
\p\zeta^J 
\right. 
\eq
\[
\left.
- \frac{1}{\phi}(\p\tilde\zeta_I + \Re {\cal N}_{IK}\p\zeta^K
)\Im {\cal N}^{IJ}(\p\tilde\zeta_J + \Re {\cal N}_{J L}\zeta^L)\right\}
\]

The coset is again given by $\frac{SU(1, N_V + 2)}{U(1)\times SU(N_V + 2)}$.
The structure of the coset metric is as given before for the non-axionic
case, but in terms of the shifted coordinates $S$ and $C$ which are 
now more complicated. They are given by,
\bea2057
S &=& \phi + \frac{1}{2} \zeta^I \Im{\cal N}_{IJ}\zeta^J + \frac{1}{2}
(\tilde\zeta_I + \Re {\cal N}_{IK}\tilde\z^K)\Im {\cal N}^{IJ}
(\tilde\z_J + \Re {\cal N}_{JL}\z^L) + i\tilde\phi \\
C_I &=& -\Im {\cal N}_{IJ} \z^J + i(\tilde\z_I + \Re {\cal N}_{IJ}\z^J) \\
\nonumber
\eeqa

In order to explore the quaternionic structure in the double extreme case,
one again has to introduce the complex 1-forms as,
\bea2058
E^A &=&e^{(\wt K - K)/2} P_I^A(N^{-1})^{IJ} d C_J \\
u &=& 2 e^{(\wt K + K)/2} z^I d C_I \\
v &=& e^{\wt K} (d S' - (C + \bar C)_I \Im{\cal N}^{IJ} d C_J \\
\nonumber
\eeqa
where $P_I^A$ is a $N_V\times (N_V + 1)$ matrix. 
The connection $\Omega$ can be written down from the covariantly constancy 
condition of the quaternionic vielbein V, {\it i.e.} 
$(d + \Omega)V = 0$. The form of $\Omega$ is given by,
\be2059
\Omega = p\otimes{\bf 1}_{2 N_V + 1} + {\bf 1}_2 \left( \begin{array}
{cc}
q & t \\
- t^{\dag} & - q^T \\
\end{array}\right)
\eq
where,
\be2060
p = \left ( \begin{array}{cc}
\frac{1}{4}(v - \bar v) & - u \\
\bar u & -\frac{1}{4}(v - \bar v) \\
\end{array} \right)
\eq
\be2070
q = \left ( \begin{array}{cc}
- \frac{3}{4}(v - \bar v) & \bar E^B \\
- E^A & -\frac{1}{4} (v - \bar v)
\end{array} \right)
\eq
\be2710
t = \left (\begin{array}{cc}
0 & 0 \\
0 & - \frac{{\bar f}_{ABC} {\bar E}^C}{4 \bar z N z}
\end{array} \right)
\eq
The curvature 2-form 
has $SU(2) \otimes Sp(2 N_V + 2)$ holonomy and can be explicitly 
computed. We do not give the details here. 

Till now, we discussed the case of double extreme black hole solutions 
corresponding to constant moduli $z^A$  and used the c-map to go to the 
dualized version in type IIB. If $z^A$ are not constant, then there 
will be additional terms in the dualized action involving terms like 
derivative of the matrix ${\cal N}$. The corresponding c-map 
becomes more complicated, but nevertheless one can again work out the
details as before and explore the quaternionic structure. 
All the dualized solutions we have found here through a c-map will 
eventually satisfy the geodesic equation of motion as mentioned in 
the beginning. Note that, we did not have to solve these inhomogeneous 
second order equations explicitly to obtain the solution. The knowledge 
of the corresponding solution in the IIA side and the T-dualized 
version (c-map) is sufficient to write down the solution in the IIB 
theory.


\subsubsection{The T-duals of static, non-axionic solutions}

During the last years extremal black hole solutions of 
$D=4$, $N=2$ supergravity coupled to vector multiplets have been
studied extensively. Such solutions can be expressed in
terms of harmonic functions, 
which are related to the different kinds
of electric and magnetic charges that the black hole carries.
The subclass of non-axionic black holes has a particular
simple structure. In order to get a better idea
about the physics of D-instantons we will consider
in this subsection the T-duals of extreme, non-axionic
black holes.

The axion-free case can be defined on the IIA side
by restricting the scalar fields to purely imaginary
values, i.e. the real part, which has an axion-like
shift symmetry vanishes: $\Re z^{A}=0$. For simplicity
we will restrict ourselves to a proper subclass where
half of
the charges are set to zero (in general we only get
$N_{V}$ relations between the $2 N_{V}+2$ charges).
For definiteness we choose
the field strength $F^{0}_{\mu \nu}$ to be purely electric,
whereas the other field strength $F^{A}_{\mu \nu}$ are
purely magnetic. Such a non-axionic black hole carries one
electric charge $q_{0}$ and $N_{V}$ magnetic  charges $p^{A}$.
The prepotential and hence the black hole solution of a generic
type II compactification is very complicated. In order to
have a sufficently simple configuration we go to a limit
in moduli space where the prepotential reduces  
to the cubic form
$F=d_{ABC} \frac{X^{A}X^{B}X^{C}}{X^{0}}$.
The static non-axionic solution for this case
has been found in \hcite{klaus} and it depends
on $N_{V}+1$ functions $H_{0}, H^{A}$ which are harmonic
with respect to the transversal 3-radius. Using the general
formulae that are derived in section 4.3 we can get
the corresponding axion-free wormhole solution. As already
explained we have to take the harmonic functions to depend
on the transversal 4-radius $r$, $r^{2}=t^{2}+x^{2}+y^{2}+z^{2}$
in order to describe a localized $(-1)$-brane.
The string and Einstein frame metrics are:
\be5310
ds_{S}^{2} = e^{2 \varphi(r)} d{s}_{E}^{2} =
e^{2 \varphi(r)} ( dr^{2} + r^{2} d\Omega_{3}^{2})
\eq
with dilaton
\be5320
e^{2 \varphi(r)} = \sqrt{4 H_{0} d_{ABC} H^{A} H^{B} H^{C}}\,.
\eq
The remaining non-trivial scalar fields are
\be5330
z^{A} = 2 i H_{0}H^{A} e^{-2\varphi},\;\;\;
\zeta^{0} = \frac{1}{\sqrt{2}H_{0}}, \;\;\;
\p_{\mu} \wt{\zeta}_{A} = - \frac{1}{2 \sqrt{2}} 
\Im {\cal N}_{AB} \p_{\mu} H^{B} 
\eq
where the IIA gauge kinetic matrix ${\cal N}_{IJ}$ is
purely imaginary, hence does not contain $\theta$-angles.
It can explicitly be expressed in terms of harmonic functions
as 
\be5340
{\cal N}_{00} = 2 i \, H_{0}^{2} \, e^{-2 \varphi}\ ,\;\;\;
{\cal N}_{0A} = {\cal N}_{A0} = 0\ ,\;\;\;
{\cal N}_{AB} = 4 \, g_{AB} \, {\cal N}_{00} = 8i \, H_{0}^{2} \,
e^{-2 \varphi} \, g_{AB} \,,
\eq 
where $g_{AB}$ is the IIA vector multiplet metric.

In order to describe a single $(-1)$-brane the harmonic 
functions have to be taken to depend on the overall radius:
\be5350
H_{0} = h_{0} + \frac{q_{0}}{r^{2}},\;\;\;
H^{A} = h^{A} + \frac{p^{A}}{r^{2}}
\eq
For $r \rightarrow \infty$ the string frame metric 
becomes flat:
\be5355
e^{2 \varphi} \rightarrow e^{2 \varphi_{\infty}}=
\sqrt{4 h_0 d_{ABC} h^A h^B h^C}
\eq
On the IIA side one has to put the constraint
\be5360
4 h_{0} d_{ABC} h^{A} h^{B} h^{C} = 1
\eq
in order to 
ensure that the black hole metric is asymptotic to the standard
Minkowski metric for $r \rightarrow \infty$.

On the IIA side the Bekenstein-Hawking entropy $S_{BH}$, the
minimized central charge $Z_{min}$ and area $A$ of the event
horizon can be expressed in terms of the charges
\be5370
\frac{1}{\pi} S_{BH} = \frac{1}{4 \pi} A =
|Z_{min}|^{2} = \sqrt{4 q_{0} d_{ABC} p^{A}p^{B}p^{C}}
\eq
It is natural to expect that this quantity will also play an
important role on the IIB side. When exploring the behaviour
for $r\rightarrow 0$ one indeed finds that $Z_{min}$ naturally 
appears. One can easily verify that the metric (\href{5310})
is invariant under the transformation 
\be5375
r \rightarrow \frac{|Z_{min}|}{r}\ ,\;\;\;
h_0 \rightarrow  \frac{q_0}{|Z_{min}|}\ , \;\;\;
q_0 \rightarrow h_0 |Z_{min}| \ ,
h^A \rightarrow \frac{p^A}{|Z_{min}|} \ ,
p^A \rightarrow h^A |Z_{min}| \ .
\eq 
Note that one has to transform charges into asymptotic moduli and
that the minimized central charge replaces the charge appearing
in the formula (\href{2017}) for the D-instanton of 
dilaton supergravity.

Let us now specialize to the double extreme case that we
already discussed from the IIB side in section 4.2.1.
This case is defined by setting the scalars $z^{A}$
to constant values, which implies that the asymptotics
of the harmonic functions is fixed by the charges, i.e.
$h_{0}$ and $h^{A}$ can be expressed in terms of $q_{0}$
and $p^{A}$:
\be5390
h_{0} = \frac{q_{0}}{|Z_{min}|}, \;\;\;
h^{A} = \frac{p^{A}}{|Z_{min}|}\,.
\eq
Note that this is consistent with (\href{5375}) because
it just puts the asymptotic moduli $h_0,h^A$ to their fixed points
under (\href{5375}). In other words: The IIB fixed point
values of the moduli, i.e. fixed points under the space inversion,
are the same as attractor fixed points in IIA which guarantee 
full supersymmetry on the event horizon.
The scalars $z^{A}$ take the constant values 
$z^{A}= i\frac{2 q_{0}p^{A}}{\sqrt{4 q_{0}d_{ABC}p^{A}p^{B}p^{C}}}$ 
and the dilaton is
\be5400
e^{2 \varphi} = \left(1 + \frac{|Z_{min}|}{r^{2}} \right)^{2}
\eq
This is precisely what we got for 
the simple D-instanton solution with
$e^{\varphi_{\infty}}=1$ and the original $q_{0}$
replaced by $|Z_{min}|= (4q_{0}d_{ABC}p^{A}p^{B}p^{C})^{1/4}$.\\
The reason that we get $e^{\varphi_{\infty}}=1$ is that
we started with a black hole which was asymptotically
flat thanks to the constraint (\href{5360}).

Now the solution has the simple inversion 
symmetry 
\be5410
r \rightarrow \frac{|Z_{min}|}{r}\,,
\eq
because the moduli are at their fixed point values.

The neck of the wormhole is localized at the selfdual radius
$r=|Z_{min}|^{1/2}$, i.e. the IIA central charge indeed 
characterizes the wormhole geometry.

\subsubsection*{Non-axionic Calabi-Yau black holes and 
D-instantons}

The non-axionic black hole and D-instanton solutions
that we discuss in this section are similar to
the four-dimensional configurations that we got in section 2
by toroidal compactification of the ten-dimensional
$(0,4,4,4)$ solution of IIA and the $(-1,3,3,3)$ solution of
IIB. We will now explain this in some more detail. 

Let us recall the relation between ten-dimensional and
four-dimensional theories in generic terms, without specifying
at the moment whether the Ricci-flat manifold we compactify
on is a torus or a Calabi-Yau threefold. For definitenes
and simplicity we will assume that we study axion-free solutions
that carry one electric and several magnetic charges. This
is the case for all the solutions discussed in sections 2 and
in sections 4.1 and 4.2.

The relation between ten- and four-dimensional R-R gauge fields
is provided by decomposing the various R-R $p$-forms
into a four-dimensional space-time and a six-dimensional
internal part. Massless four-dimensional gauge fields
are obtained if the internal part is a harmonic differential
form. Thus the number of massless four-dimensional gauge fields
is determined by topological data, namely the Betti numbers of
the internal manifold. On the other hand the R-R charges of
the four-dimensional theory come from the charges carried
by the various D-$p$-branes of the ten-dimensional theory. 
Upon compactifications the branes are wrapped around 
homology cycles of the internal manifold.
Poincare duality between non-trivial cycles and harmonic
differential forms ensures that the number of charges 
equals the number of gauge fields in four dimensions, which
is of course needed for consistency. Finally all parameters
describing marginal deformations of the internal metric 
and other internal fields appear as moduli scalars in the
four-dimensional theory.

Let us for definiteness consider a configuration which
consists of 3-branes  which are wrapped around 4-cycles
of the internal manifold together with $(-1)$ branes
located at points (0-cycles). The simplest case, with the
internal manifold being a torus was discussed in section 2.
Note that the complete worldvolume, including the world-sheet 
time direction is wrapped, so that the configuration is 
an instanton from the four-dimensional point of view.
Taking $q_{0}$ $(-1)$-branes together with $p^{A}$ 
3-branes in the $A$-th primitive homology class, one
gets a four-dimensional solution with electric charge
$q_{0}$ and magnetic charges $p^{A}$. In the case of
a torus, there are three primitive 4-cycles, 
for which we have specified the explicit representatives in section 2
(in general the Betti numbers of a $d$-dimensional torus
are just $b_p = d! / ( p! (d-p)! )$).
Therefore the solution could be expressed in terms
of four harmonic functions $H_{0}$, $H^{1}$, $H^{2}$ and
$H^{3}$. 

General static black hole solutions of four-dimensional $N=2$
supergravity can be expressed in terms of harmonic functions
by using special geometry \hcite{sabra}. 
For a Calabi-Yau compactification the 
number of harmonic functions
$H^{A}$ will depend on the number of primitive
4-cycles, which is $b_4 = h_{1,1}$. 
Passing from cycles to harmonic forms by
Poincare duality we can analyse the dimensional reduction of
the gauge fields. {From} the material recalled in section 3.2
it is obvious that all non-trivial ten-dimensional 
R-R gauge fields become four-dimensional R-R scalars
which sit in hyper multiplets. One can also look what happens
to the NS-NS scalars that describe deformations of the
internal metric. 
In the toroidal case we found in section 2 that the solution leads 
to three non-trivial moduli  
$T_{1}$, $T_{2}$ and $T_{3}$, which are related to the size of
certain 2- and 4-cycles of the internal torus.
For the Calabi-Yau case we again find that the non-trivial moduli
appearing in the solution are related to 2- and 4-cycles and using
the material reviewed in section 3 one sees that they sit in
hyper multiplets. 
Moreover one
can check that all the fields in the four-dimensional
gravity and vector multiplets descend from ten-dimensional
fields that are trivial in a solution based on $(-1)$-branes
and 3-branes. Thus all fields that are non-trivial 
in the four-dimensional solution sit in hyper multiplets.

Let us finally consider the limit in which the Calabi-Yau
compactification is described by a cubic prepotential
on the IIA side. We found in the last subsection that
the string frame geometry of the IIB instanton depends on the 
combination $H_{0}d_{ABC}H^{A}H^{B}H^{C}$ of harmonic
functions. To explain the relation with the 
combination $H_{0}H^{1}H^{2}H^{3}$ that we found in the
toroidal case, we have to recall that the parameters 
$d_{ABC}$ in the case of a Calabi-Yau compactification
are related to the triple-intersection numbers 
$C_{ABC}$ by $d_{ABC} = \frac{1}{3!}C_{ABC}$. 
A torus has three primitive homology classes of 4-cycles
and its intersection numbers are $C_{ABC}=1$ if $A,B,C$
are all different and $C_{ABC}=0$ otherwise. Thus, dependence of
the dilaton on harmonic functions is the same 
for both torus and
Calabi-Yau compactification (in the limit of a cubic
prepotential).

\subsection{Dualizing the general IIA solution}

In this section, we will T-dualize ($c$-map) the general IIA
solution. As a result we will find a IIB instanton solution.

{\bf The solution}

For our type IIA solution we assume that the scalar fields in the 
hyper multiplets are trivial,
the corresponding action is given in eq.\ (\href{0010}).
In order to dualize the fields we need an isometry, which is in our
case the Euclidean time.
Then the supersymmetric solution is given by \hcite{be/lu} 
\be1020
\ba{c}
ds^2 =  e^{2U} (dt + \omega_m dx^m)^2 + e^{-2 U} dx^m dx^m \ ,\\
F^I_{mn} = {1 \over 2} \epsilon_{mnp} \partial_p \tilde{H}^I
\quad , \quad G_{I\, mn} = {1 \over 2} \epsilon_{mnp} \partial_p H_I \ , \\ 
z^A = {X^A \over X^0} \quad , \quad q^u = \mbox{const.}
\ea
\ee 
with
\be1030
\ba{rcl}
G_{I\, \mu\nu} & = &\Re {\cal N}_{IJ} F^{J}_{\mu\nu} - 
 \Im {\cal N}_{IJ} {^{\star}}F^J_{\mu\nu} \ , \\
e^{-2U}=e^{-K}  & \equiv & i (\bar{X}^I F_I- X^I \bar{F}_I) \ , \\
{1 \over 2} e^{2U} \epsilon_{mnp} \partial_n \omega_p = Q_m & \equiv &
 {1 \over 2} e^{K}( \bar{F}_I \partial_m X^I - \bar{X}^I \partial_m F_I
 + c.c. ) \\
 &=& { 1 \over 2} e^{2U} ( H_I \partial_m \tilde{H}^I - \tilde{H}^I
\partial_m H_I) 
\ea
\ee
where the holomorphic section $(X^I , F_I)$ is constrained by
\be1040
i (X^I - \bar{X}^I ) = \tilde{H}^I(x^{\mu}) \qquad , \qquad 
i (F_I -\bar{F}_I) = H_I(x^{\mu})  \ .
\ee
Since the symplectic vector $( \tilde H^I , H_I )$ is introduced
via the gauge fields, these equations can be seen as a constraint
between the holomorphic section $(X^I , F_I)$ and the gauge field
section $( F^I_{\mu\nu}, G_{\mu\nu\, I})$.

In order to find a non-trivial Killing spinor
one has to require an integrability condition 
\be1042
\partial_{[m} Q_{n]} =0 \qquad \mbox{or:} \quad 
\partial_{[n} \tilde H^I \partial_{m]} H_J = 0 \ .
\ee
As a consequence, we can express $\omega_m$ by a further harmonic 
function $\hat H$
\be1044
\epsilon_{mnp} \partial_n \omega_p = ( H_I \partial_m \tilde{H}^I - 
\tilde{H}^I \partial_m H_I) ={1 \over 2} \partial_m \hat H \ .
\ee
For these fields, the supersymmetry variations vanish 
\be1043
\ba{rcl}
\delta\,\psi_{\alpha\mu} &=& \nabla_\mu \epsilon_\alpha -
\frac{1}{4} T^-_{\rho\sigma} \gamma^{\rho} \gamma^{\sigma}
\, \gamma _\mu \, \varepsilon_{\alpha\beta}\epsilon^\beta \, 
\label{grtrans}=0 ,\\
\delta\lambda^{A\alpha} & = & i \, \gamma^\mu \partial_\mu
z^A \epsilon^\alpha + G^{-A}_{\rho\sigma}\gamma^{\rho}\gamma^{\sigma}
\varepsilon ^{\alpha\beta} \epsilon_\beta = 0
\ea
\ee
with
\be1059
\nabla_\mu \epsilon_\alpha=(\partial_\mu-{1\over4}w^{ab}_\mu\gamma_a 
\gamma_b + {i\over 2}Q_\mu)\epsilon_\alpha 
\ee 
where $w^{ab}_\mu$ is the spin connection. 
If in addition the functions $\left(\tilde{H}^I(x^{\mu}) , H_I(x^{\mu})
\right)$ are harmonic, then the equations of motion are also fulfilled.

\bigskip

{\bf T-duality}

Next we dualize the action (\href{0010}) by compactifying over
an radius $R$ and decompactifying over $1/R$ (as usually for 
T-duality, see also \hcite{BerDeR}). 
The only subtlety in this procedure occurs from
the gauge field part. For the curvature and scalar part
the standard rules give (including surface terms and in order to 
avoid confusions, we will often supress all space time indices.)
\be1050
\ba{l}
\int d^{4}x \sqrt{G} \left[ R - 2 g_{A\bar{B}} 
 \partial z^A \, \partial \bar{z}^B\right] \rightarrow  \hfill \\
\qquad \int  d^4 x \sqrt{G} \, e^{- 2\varphi} \left[R + 4 (\partial 
\varphi)^2
- 2 e^{2 \varphi} {1 \over \sqrt{g}} \partial( e^{-\varphi} \partial 
e^{-\varphi})
- {1 \over 12} \hat {\cal H}^2
-2 g_{A\bar{B}}  \partial z^A \, \partial \bar{z}^B \right] \ .
\ea
\ee
The new fields are
\be1060
\ba{l}
ds^2 = e^{2 \varphi} dx^{\mu} dx^{\mu} \quad , \quad e^{2\varphi} = 
e^{-2U} \ ,\\
\hat {\cal H}_{0mn} = \partial_m \omega_n - \partial_n \omega_m \ .
\ea
\ee 
As a next step, we have to consider the gauge field part. First one
has to go to three dimensions. Using standard formulae for the
dimensional reduction (see e.g. \hcite{ma/sc})
we obtain for the 3-d Lagrangian 
\be1070
\ba{l}
S_3 = \int d^3x \sqrt{g} \, e^{U} \left[- {1 \over 4}  
 (F^{(3)\, I} - \sqrt{2} \, \zeta^I d\omega) \,\Im {\cal N}_{IJ}\,
(F^{(3)\, J} - \sqrt{2} \, \zeta^J d\omega) - \right. \\ \qquad \left. -
e^{-2 U} \, \Im {\cal N}_{IJ}\, \partial \zeta^I \partial \zeta^J + 
 \sqrt{2} \, e^{-U}\, \Re {\cal N}_{IJ}\, \partial \zeta^I {^{\star}}
( F^{(3)\, J} - \sqrt{2}\, \zeta^J d\omega) \right]
\ea
\ee
with
\be1080
A^I_0 = \sqrt{2} \zeta^I \quad , \quad
 F^{(3)\, I }_{mn} = {1 \over 2} \epsilon_{mnp} \partial_p \tilde H^I 
\quad , \quad
(d\omega)_{mn} = \partial_m \omega_n - \partial_n \omega_m \ .
\ee
After decompactification over the inverse radius, i.e.\ $G_{00}
\rightarrow 1/G_{00}$  and adding the curvature part (\href{1050})
we find for the 4-d theory 
\be1090
\ba{l}
S_4 = \int  d^4 x \sqrt{G} \, e^{- 2\varphi} \left[ R + 4 (\partial 
\varphi)^2  - {1 \over 12} \hat{\cal H}^2
- 2 e^{2 \varphi} {1 \over \sqrt{g}} \partial( e^{-\varphi} \partial 
e^{-\varphi})
-2 g_{A\bar{B}}  \partial z^A \, \partial \bar{z}^B - 
\right. \\ \qquad \left.
- e^{2 \varphi}\, \Im {\cal N}_{IJ}\,  \partial \zeta^I \partial \zeta^J
 - {1 \over 12} e^{2 \varphi} ({\cal H}^I - \sqrt{2} \, 
\zeta^I \hat{{\cal H}}) 
 \,\Im {\cal N}_{IJ}\, ({\cal H}^{J} - \sqrt{2} \,
 \zeta^J \hat{{\cal H}}) + \right. \\ \qquad \left.+
 \sqrt{2} \,e^{2 \varphi} \,\Re {\cal N}_{IJ}\, 
\partial \zeta^I\, {^{\star}}
( {\cal H}^{J} -  \sqrt{2} \,\zeta^J \hat{{\cal H}}) \right]
\ea
\ee
where the new antisymmetric tensors are
\be1100
{\cal H}^I_{0mn} = F^{(3)\, I}_{mn} \quad , \quad ({^{\star}{\cal H}})^{\mu}
= {1 \over 6 \, \sqrt{G}} \epsilon^{\mu\nu\rho\lambda} 
{\cal H}_{\nu\rho\lambda} \ .
\ee
This is the type IIB action in the string frame. We go to the Einstein 
frame with the relation,
\be1110
g^E_{\mu\nu} = e^{-2 \varphi} g_{\mu\nu} = \delta_{\mu\nu}  \ .
\ee
Again, including all surface terms one obtains
\be1120
\ba{l}
S_4^E = \int d^4x \left[ - 4 \, \partial^2 \varphi - 2 (\partial \varphi)^2
-2 g_{A\bar{B}}  \partial z^A \, \partial \bar{z}^B - 
e^{2 \varphi} \Im {\cal N}_{IJ}  \partial \zeta^I \partial \zeta^J - 
\right. \\ \qquad \left.
- {1 \over 12} e^{-4 \varphi}  \hat{{\cal H}} - {1 \over 12} e^{-2 \varphi} 
({\cal H}^I -  \sqrt{2}\, \zeta^I \hat{{\cal H}})\, \Im {\cal N}_{IJ}\, 
({\cal H}^{J} - \sqrt{2}\, \zeta^J \hat{{\cal H}}) + \right. \\ \qquad 
\left. +  \sqrt{2}\,\Re {\cal N}_{IJ}\, \partial \zeta^I\, {^{\star}}
( {\cal H}^{J} -  \sqrt{2}\, \zeta^J \hat{{\cal H}}) \right]
\ea
\ee
Using our supersymmetric configuration (\href{1020}),
the IIB antisymmetric tensors can be written in a covariant way, 
namely,
\be1130
{\cal H}^{\mu\nu\rho\, I} = {1 \over 2}\,
\epsilon^{\mu\nu\rho\lambda} \partial_{\lambda}
H^I \quad , \quad {\cal H}^{\mu\nu\rho} = {1 \over 2} \,
\epsilon^{\mu\nu\rho\lambda} \partial_{\lambda} \hat H
\ee
where $\hat H$ is given in (\href{1044}). Like on the IIA side, the
Bianchi identities are only fulfilled if all $H's$ are harmonic.  This
formula already suggests that on the type IIB side we can relax the
isometry along the $t$-direction by allowing harmonic functions 
depending on 
all four coordinates. As in 10 dimsensions, this 
localizes the solution in all transversal directions, i.e.\ for
the instanton, also in the time direction. If the harmonic
functions depend only on three coordinates we recover the
expressions given in (\href{1060}) and (\href{1100}).

As next step we dualize the antisymmetric tensors $({\cal H}^I , 
\hat {\cal H})$ into further scalar
fields, where we  use the same notation as introduced in section 3.
In doing this, we have to first add a Lagrange multiplier term,
\be1140
\ba{l}
\delta S = - \int \left[ (\tilde \phi - \zeta^I \tilde \zeta_I)\, \partial 
 {^{\star}\hat{{\cal H}}} + \sqrt{2}\, \tilde \zeta_I \partial {^{\star}
 {\cal H}^I} \right] \\ \quad
 =  \int \left[ \partial(\tilde \phi - \zeta^I \tilde \zeta_I)
 {^{\star}\hat{{\cal H}}} + \sqrt{2}\, \partial \tilde \zeta_I 
 {^{\star}{\cal H}^I} \right]  - \int \partial \left[ (\tilde \phi - 
\zeta^I \tilde \zeta_I) {^{\star}\hat{{\cal H}}}  + \sqrt{2}\,  
\tilde \zeta_I {^{\star}{\cal H}^I} \right] \ .
\ea
\ee
to the previous action. Note that the second term is a further total 
derivative term.
Integrating $\tilde \phi$ and $\tilde \zeta_I$ yields the
Bianchi identities. On the other hand taking the variations
with respect to the torsions yields
\be1510
{\cal H}^{J} - \sqrt{2}\,\zeta^J \hat {\cal H} = 
- \sqrt{2}\, e^{2 \varphi}
\Im {\cal N}^{JI} \, {^{\star}(\partial \tilde \zeta_I + 
\Re {\cal N}_{IL}\, \partial \zeta^L )} \ .
\ee
After inserting this field into the action and variing 
$\hat {\cal H}$ we find 
\be1520
\hat {\cal H} = - e^{4 \varphi} \, {^{\star}( \partial \tilde \phi 
+ \zeta^I \partial \zeta_I - \tilde \zeta \partial \zeta)} \ .
\ee
So, we end up with the 4-d dualized Einstein action
\be1530
\ba{l}
S_4^E = \int d^4x \left[ -2 (\partial \varphi)^2
 -2 g_{A\bar{B}}  \partial z^A \, \partial \bar{z}^B -   e^{2 \varphi}   
 \partial \zeta^I \, \Im {\cal N}_{IJ}\, \partial \zeta^J + 
 \right. \\ \qquad \left. 
 + e^{2 \varphi} (\Re {\cal N}_{IL} \partial \zeta^L +  
 \, \partial \tilde \zeta_I)
 \, \Im {\cal N}^{IJ}\, (\Re {\cal N}_{JK} \partial 
 \zeta^K +  \, \partial \tilde \zeta_J) 
+  \right. \\ \qquad \left. + {1 \over 2} e^{4 \varphi} 
(\partial \tilde \phi +\zeta^I \partial \tilde \zeta_I - 
\tilde \zeta_I \partial \zeta_I )^2\right] + S_{\partial M} \ .
\ea
\ee
The surface terms are
\be1540
\ba{l}
S_{\partial M} = - \int d^4x \, \partial \left[ 4 \, \partial \varphi +
2 \, e^{2 \varphi}\, \tilde \zeta_I \, \Im{\cal N}^{IJ} \,
( \Re {\cal N}_{IL} \partial \zeta^L + \, \partial \tilde \zeta_I)
+ \right. \\ \left. \qquad +
\, e^{4 \varphi} \,( \tilde \phi + \tilde \zeta_I \zeta^I)
(\partial \tilde \phi + \zeta^I \partial \tilde \zeta_I
- \zeta^I \tilde \partial \zeta_I) \right]  
\ea
\ee
Inserting the expressions for the antisymmetric tensors as given in
(\href{1130}), we can express the IIB solution in terms of the IIA
fields defined in (\href{1020}) and (\href{1030})
\be1550
\ba{l}
e^{2 \varphi} = e^{-2 U} \qquad , \qquad \sqrt{2} \, \zeta^I = A^I_0 \qquad ,
\qquad z^A = {X^A \over X^0}\ , \\
2\, \partial \tilde \phi = - e^{-4 \varphi} \partial \hat H + \sqrt{2} \,
\tilde \zeta_I \partial A_0^I - \sqrt{2} \, A_0^I \partial \tilde \zeta_I 
\ ,\\
\sqrt{2}\, \partial \tilde \zeta_I = - {1\over 2} e^{-2 \varphi} \,
\Im {\cal N}_{IJ} \, ( \partial H^J - A_0^J \partial \hat H) -
\Re {\cal N}_{IJ} \partial A_0^J \ , 
\ea
\ee
where $\hat H$ is defined in (\href{1044}). Since $\tilde \phi$ and
$\tilde \zeta$ are scalars that appear via dualization of
antisymmetric tensors, we have only an expression for their
derivatives.  The integrability constraint (\href{1042}) ensures
that one can integrate these equations.
Again, off-shell it is supersymmetric for any functions
($H^I , H_I$), but on-shell they have to be harmonic, see Bianchi
identity for the torsion in (\href{1130}).


\section{Conclusions}

In this paper we discussed the T-duality between the extremal
black hole type  solutions
in compactified four-dimensional IIA strings and
and  the D-instanton solutions in IIB superstrings compactified on the
same internal  Calabi-Yau space. In the context of the effective 
four-dimensional $N=2$ supergravity, the T-duality precisely acts like
the c-map which exchanges the special K\"ahler moduli space of the
type IIA vector multiplets with the quaternionic moduli space of the
IIB hyper multiplets. Hence, the IIA black holes are the BPS solutions
of the special K\"ahler Lagrangian, whereas the D-instantons arise as
the solutions of the dual quaternionic $\sigma$-model.
This scenario opens room for several interesting discussions which are
worth exploring further. 
For example we find that our wormhole solutions connect two flat regions,
where one of them is strongly coupled and both regions are related by a 
symmetry of the solution.  It would be important to clarify
the precise meaning of the IIA black hole entropy in the
context of the IIB D-instantons. 
We observed that for the double-extreme case (with extremized 
central charge) the black
hole entropy
becomes the self-dual radius in the D-instanton wormhole geometry; so we
might ask what precisely happens at the self-dual wormhole radius. 
And the relation 
of the self-dual radius to the microscopic D-brane counting is still
mysterious. Finally the D-instanton solutions we found will be relevant
for the computation of stringy instanton effects along the lines of
\cite{GreGut,GreenVanhove}.

\section*{Acknowledgements}

We would like to thank B. Ovrut and W. Sabra for discussions.
Moreover we would like to thank B. Ovrut and D. Waldram for 
pointing out an error in the first version of our paper.
Work of K.B. is supported by DFG. Work of T.M. is supported by DFG
and by the European Commission programme
ERBFMRX-CT96-0045. Work of S.M. is supported by Alexander von Humboldt 
Foundation.





\end{document}